\documentclass[%
 reprint,
 amsmath,amssymb,
 aps,nofootinbib,
superscriptaddress]{revtex4-1}
\usepackage{tablefootnote}
\usepackage{dsfont}
\usepackage{graphicx}
\usepackage{dcolumn}
\usepackage{bm}
\usepackage{color}
\usepackage{epsfig}
\usepackage{hyperref}
\usepackage{natbib}
\usepackage{float}
\usepackage{footmisc}
\usepackage{aas_macros}
\usepackage{slashed}
\usepackage{comment}
\usepackage{braket}
\usepackage{soul}
\usepackage{fontawesome5}
\definecolor{orcidlogocol}{rgb}{0.65, 0.807, 0.223}
\newcommand{\orcid}[1]{\,\href{https://orcid.org/#1}{\textcolor{orcidlogocol}{\footnotesize\faOrcid}}}

\definecolor{lightred}{rgb}{1,0.5,0.5}
\definecolor{lightgreen}{rgb}{0.5,1,0.5}
\definecolor{lightblue}{rgb}{0.5,0.5,1}
\definecolor{lightcyan}{rgb}{0.5,0.75,0.75}
\definecolor{lightmagenta}{rgb}{0.75,0.5,0.75}
\definecolor{customgreen}{rgb}{0.494,1,0.502}

\newcommand{\keV}{\mathinner{\mathrm{keV}}}
\newcommand{\MeV}{\mathinner{\mathrm{MeV}}}
\newcommand{\GeV}{\mathinner{\mathrm{GeV}}}
\newcommand{\TeV}{\mathinner{\mathrm{TeV}}}

\begin{document}

\title{Indirect Searches for Ultraheavy Dark Matter in the Time Domain}

\author{David E. Kaplan}
\email{david.kaplan@jhu.edu}
\affiliation{The William H.~Miller III Department of Physics and Astronomy, The Johns Hopkins University, Baltimore, Maryland, 21218, USA}
\author{Xuheng Luo}
\email{xluo26@jhu.edu}
\affiliation{The William H.~Miller III Department of Physics and Astronomy, The Johns Hopkins University, Baltimore, Maryland, 21218, USA}
\author{Ngan H.~Nguyen}
\email{nnguye53@jhu.edu}
\affiliation{The William H.~Miller III Department of Physics and Astronomy, The Johns Hopkins University, Baltimore, Maryland, 21218, USA}
\author{Surjeet Rajendran}
\email{srajend4@jhu.edu}

\affiliation{The William H.~Miller III Department of Physics and Astronomy, The Johns Hopkins University, Baltimore, Maryland, 21218, USA}
\author{Erwin H.~Tanin}
\email{ehtanin@stanford.edu}
\affiliation{Stanford Institute for Theoretical Physics, Stanford University, Stanford, California, 94305, USA}

\begin{abstract}
Dark matter may exist today in the form of ultraheavy composite bound states. Collisions between such dark matter states can release intense bursts of radiation that includes gamma-rays among the final products. Thus, indirect-detection signals of dark matter may include unconventional gamma-ray bursts. Such bursts may have been missed not necessarily because of their low arriving gamma-ray fluxes, but rather their briefness and rareness. We point out that intense bursts whose non-detection thus far are due to the latter can be detected in the near future with existing and planned facilities. In particular, we propose that, with slight experimental adjustments and suitable data analyses, imaging atmospheric Cherenkov telescopes (IACTs) and Pulsed All-sky Near-infrared and Optical Search for Extra-Terrestrial Intelligence (PANOSETI) are promising tools for detecting such rare, brief, but intense bursts. We also show that if we assume these bursts originate from collisions of dark matter states, IACTs and PANOSETI can probe a large dark matter parameter space beyond existing limits. Additionally, we  present a concrete model of dark matter that produces bursts potentially detectable in these instruments.
\end{abstract}

\maketitle
\tableofcontents
\newpage
\section{Introduction}
Numerous programs have been conducted to search for possible electromagnetic signals of dark matter decay and annihilation \cite{Hooper:2018kfv,Slatyer:2017sev,Safdi:2022xkm,Gaskins:2016cha}. Such indirect dark matter searches have spanned broadly across the electromagnetic spectrum, covering many orders of magnitude in the frequency domain. Due to strong emphasis on minimal and popular dark matter models such as weakly interacting massive particles \cite{Jungman:1995df,Conrad:2014tla,Roszkowski:2017nbc} and axion-like particles \cite{Caputo:2024oqc,AxionLimits}, the majority of the searches have been geared toward persistent signals. On the other hand, the Standard Model universe hosts a large assortment of astrophysical objects that produce a great diversity of transient signals with rich profiles in the time domain: flares from pulsar wind nebulae, jets from microquasars, outbursts from cataclysmic variable stars, etc \cite{Rieger:2019nna,2012IAUS..285.....G,van2012relativistic}. This motivates us to expand the discovery space of indirect dark matter detection by systematically covering not only across the electromagnetic spectrum (frequency domain) but also over a broad range of temporal structures (time domain).

As a start, we consider gamma-ray transient signals, which are well parametrized by their arriving fluxes, occurrence rates, and time durations. A thorough search for such signals should aim not only at achieving sensitivity to the smallest fluxes but also at covering broad ranges of occurrence rates and time durations. Even bursts that release a huge amount of energy and arrive with very high fluxes can be missed if they are rare and very brief. Catching signals that are off most of the time and appear at unpredictable locations on the sky requires detectors with high exposure: large field of view (FoV) and high duty cycle. Detecting short-duration events poses its own challenges due to fundamental hardware limits on the sampling rate of a detector. Additional limits may arise from practical implementation details of the detectors such as their trigger algorithms and data transmission speeds.

Fermi-LAT, the current-generation space-based gamma-ray detector, covers $\mathcal{O}(1)$ of the sky with $\mathcal{O}(1)$ duty cycle, but it is essentially blind to gamma-ray bursts with durations less than about ten microseconds due to its electronic hardware limitations \cite{Atwood_2009,1993ApJ...404..673S}. Besides that, ground-based imaging atmospheric Cherenkov telescopes (IACTs) \cite{2023hxga.book..144E,Prandini:2022wcb,Bose:2021oiq,2011MNRAS.416.3075L}, while, in principle, have nanosecond time resolutions, have not been utilized to their full potential when it comes to searching for $\lesssim 10\,\mu\text{s}$ duration bursts, as we will explain further in Section~\ref{section:detectiontechniques}. Moreover, IACTs have a typical FoV of $\sim 10^{-3}\text{ sr}$ and a typical duty cycle of $10\%$, which means in a decade-long observational program they are only sensitive to galactic bursts happening frequently at the rate of $\gtrsim 10^4\text{ event/year}$ on the entire sky, assuming any burst occurring within the FoV is detectable. We have thus identified an interesting observational target to aim for, namely \textit{ultrashort} ($\lesssim 10\,\mu\text{s}$ duration) gamma-ray bursts. Since this observational window has been underexplored thus far, there is a potential for discovery when our detectors become sensitive to it. The goal of this paper is to study the prospects for probing this observational space in the near future using existing and planned experimental facilities.

To further motivate exploratory searches for such ultrashort burst signals, let us recount how discoveries in time-domain astronomy occurred historically. Many of these discoveries, including those of gamma-ray bursts and pulsars, happened unexpectedly as new instruments turn on and enable us to access new observational windows \cite{2009arXiv0908.2784F}. Notably, fast radio bursts occur frequently over the whole sky at the rate of $\sim 10^3$ events/day \cite{CHIMEFRB:2021srp}, and yet we missed them initially because historically radio transient observations were only done in a targeted way on known variable objects such as active galactic nuclei and x-ray binaries. Multiple fast radio burst signals were eventually found through new analyses of the archival data of survey radio telescopes. These examples indicate that a large diversity of electromagnetic signals may have been missed simply because we have not optimized existing detectors to search for them or performed appropriate data analyses to reveal their existence.

While primarily our motivation is to probe an unexplored regime in the space of observables that is within technological reach, it is interesting to speculate on possible sources of the class of signal for which we seek. Hence, in this paper we also construct and study a concrete model of dark matter that produces ultrashort gamma-ray bursts.
An extremely short timescale and a large energy release imply a phenomenon involving a spatially-small object with extraordinary density. Given that the light crossing time of a neutron star, the most compact Standard Model object known, is $\sim 10\,\mu\text{s}$, a sub-$10\,\mu\text{s}$ burst that is strong enough to be detectable may require a beyond the Standard Model source \cite{scarglepoint}. One possibility is that the dark matter exists in the form of highly dense composite objects, dubbed \textit{dark blobs} in this paper, that produce gamma-ray bursts when two of them collide \cite{Gresham:2018anj,Bai:2018dxf,Hardy:2014mqa}.\footnote{Another potential source of short gamma-ray burst is perhaps the runaway evaporation of black holes with mass $\ll 10^{15}\text{ g}$ whose lifetimes are much shorter than the age of the universe. Such black holes may exist today if they form late \cite{Picker:2023ybp,Boluna:2023jlo}. Accounting only for Standard Model particle spectrum, a sub-$10\,\mu\text{s}$ duration burst is only achieved when the mass of the black hole is tiny, resulting in an extremely dim signal. However, if black hole evaporation proceeds differently from the standard Hawking evaporation, the duration of the explosion may be shorter for a given black hole mass. See also \cite{Crumpler:2023toy}.}  As heavy blobs easily evade \textit{direct} dark matter searches due to their rare terrestrial transits \cite{Jacobs:2014yca,Grabowska:2018lnd,Ebadi:2021cte}, currently viable dark blob parameter space is vast and still allows for strong and distinctive \textit{indirect} detection signals.

The rest of the paper is organized as follows. We explore methods for detecting ultrashort gamma ray bursts and project possible near-future sensitivities to them in Section.~\ref{section:detectiontechniques}, provide simple parametrization for ultrashort 
gamma-ray bursts from dark blob collisions and map the sensitivity projections of Section.~\ref{section:detectiontechniques} onto the parameter space of dark blobs in  Section.~\ref{section:EMsignals}, briefly discuss a simple model of dark blobs that produce ultrashort gamma-ray bursts in Section.~\ref{ss:modelsummary} (detailed further in Appendix.~\ref{appendix:modeldetails}), review existing dark blob formation mechanisms in Section.~\ref{ss:formation mechanism}, identify interesting directions for future exploration in Section.~\ref{section:discussion}, and conclude in Section~\ref{section:conclusion}. For better readability, we collect some supplementary details in Appendices. In Appendix.~\ref{appendix:wavefronttechnique}, we provide simple estimates for the sensitivities of IACTs and PANOSETI to ultrashort gamma-ray bursts based on the wavefront technique; in Appendix.~\ref{appendix:fireball} we discuss the condition for and consequences of fireball formation following a spatially and temporally concentrated injection of Standard Model particles; in Appendix.~\ref{appendix:modeldetails} we describe the details of the model summarized in Section.~\ref{ss:modelsummary}; in Appendix.~\ref{appendix:ratecalculations} we provide calculational details of the gamma-ray burst signal in the dark matter model of Section.~\ref{ss:modelsummary} and Appendix.~\ref{appendix:modeldetails}.

\section{Detecting Rare Ultrashort Gamma-ray Bursts}
\label{section:detectiontechniques}

Many orders of magnitude in the time domain below the timescale of ten microseconds remains a largely unexplored territory for gamma-ray transient searches \cite{LeBohec:2002hq,LeBohec:2005wt,Krennrich:1999tu}. If the dark matter has been emitting intermittent gamma-ray bursts with durations less than about ten microseconds, indirect searches for dark matter thus far would most certainly have missed them.

Consider as an example a scenario where the entire Galactic dark matter exists in the form of blobs, i.e. large dark matter bound states, with mass $10^{17}\text{ g}$ and radius $1\text{ cm}$. Assuming these blobs move with a typical velocity of $v_{\rm DM}=10^{-3}$ and collide with the geometrical cross section, blob collisions would occur all over the Milky Way at the rate of $\sim 1\text{ collision/year}$. When two blobs collide, suppose that a significant fraction of the blobs' mass energy $10^{17}\text{ g}\sim 10^{38}\text{ erg}$ is released in the form of gamma-rays over a short timescale, possibly set by the collision timescale $1\text{ cm}/v_{\rm DM}\sim 10\text{ ns}$. Within the brief burst duration, the luminosity of such an event can be as high $10^{45}\text{ erg/s}$, comparable to the rate at which energy is released in a supernova. If the burst originates from a collision within the Galaxy at a distance of $1\text{ kpc}$, the resulting energy fluence\footnote{Fluence here means the number of photons per unit ground area or, equivalently, the time-integrated flux of the arriving photons. Energy fluence is fluence multiplied with the (average) energy per photon.} would be $\approx 5\GeV/\text{m}^2$, which is in principle detectable since this is higher than the sensitivities of existing gamma-ray detectors.

However, the briefness and rareness of the bursts in this example scenario make them challenging to detect in practice. For such a short burst duration, the detector's pulse pile-up time and dead time become important limiting factors to consider.\footnote{Pulse \textit{pile-up time} is the time window within which arriving photons are lumped into a single event. \textit{Dead time} is the time after each event during which the detector is unable to record another event.} Moreover, these burst events might be rejected by the standard trigger systems employed in current and future gamma-ray detectors as these detectors are optimized for conventional sources, e.g. conventional $\gtrsim 1\text{ s}$-long gamma-ray bursts. As such, specialized trigger systems may be required to detect these events. Further, that the bursts occur infrequently at random positions on the sky make them extra challenging to catch with detectors that have small FoVs and low duty cycles. In this section, we discuss strategies for overcoming such challenges in detecting gamma-ray burst signals that are highly intense, but extremely \textit{brief} in duration and \textit{rare} in occurrence.

\subsection{Sub-\boldmath $10\,\mu\text{s}$ Blind Spot}

Space-based detectors, which detect transiting gamma-ray photons directly, are limited by the finite processing time it takes for electronic devices in the detectors to convert incident gamma-rays into electrical signals \cite{Atwood_2009,1993ApJ...404..673S,scarglepoint}. This electronic processing time is typically longer than a microsecond. For instance, Fermi-LAT has a pulse pile-up time of $\sim 1\,\mu\text{s}$, which means a gamma-ray burst shorter than $1\,\mu\text{s}$ will be recorded as a \textit{single} pile-up event with perceived energy given by the sum total of the energies deposited to the calorimeter onboard by two or more photons arriving within the $1\,\mu\text{s}$ window. Sub-microsecond gamma-ray bursts would appear in the data as occasional fake excesses of high-energy events from the summation of lower energy ones that do not reflect the incident gamma-ray spectrum, and it would not be possible to infer if these events were due to bursts of photons. LAT also has a dead time of $26\,\mu\text{s}$, and that means that, after each triggered event, the detector will not be able to record another event for a period of $26\,\mu\text{s}$. Thus, LAT will register more than one photon only if the burst lasts significantly longer than $26\,\mu\text{s}$.

IACTs such as HESS, MAGIC, and VERITAS avoid these issues by utilizing the atmosphere to convert incoming gamma-ray photons into optical photons, which are easier to process \cite{2023hxga.book..144E,Prandini:2022wcb,Bose:2021oiq,2011MNRAS.416.3075L}. This conversion occurs automatically as gamma-ray photons that enter the Earth's atmosphere initiate extensive showers of secondary charged particles, which emit photons in the optical range through Cherenkov radiation. IACTs detect incoming gamma-rays indirectly by collecting their Cherenkov light yields using a large aperture ($\sim 10\text{ m}$) mirror and focusing them into a pixelized camera. By analyzing the parameters (centroid position, size, shape, orientation, etc) of the digital images captured by the camera, the properties of the primary gamma-rays that initiated the showers can be inferred. Since IACTs employ fast, $\mathcal{O}(\text{ns})$ response time, photomultiplier tubes (PMTs) as the default photosensors, they have the inherent capability to probe the time domain at nanosecond or longer timescales. Moreover, compared to space-based gamma-ray detectors, IACTs usually have orders of magnitude better effective collection areas (and hence much better fluence sensitivities) because the process of air shower greatly enlarges the area of influence of the incoming gamma-ray photons to essentially the area of their showers.

While the nanosecond time resolution and large collection area of IACTs make them suitable for detecting gamma-ray bursts with ultrashort durations, IACTs unfortunately tend to have poor sky coverages \cite{2023hxga.book..144E}. For example, VERITAS covers only a $\theta_{\rm FoV}=3.5^\circ$ diameter FoV corresponding to a solid angle of $\Omega_{\rm FoV}= 2\pi\left[1-\cos\left(\theta_{\rm FoV}/2\right)\right]= 3\times 10^{-3}\text{ sr}$ at a given time, and so they would miss an event if they do not happen to be pointing at the right part of the sky. Furthermore, since IACTs must operate only during moonless nights and in good weather conditions, their duty cycles are usually low. VERITAS, for example, has only $\sim 1000\text{ hr}$ of effective observation duration per year. This reduces further the chance that a burst appears in the detector's FoV while it is collecting data.

To summarize, gamma-ray bursts in the sub-$10\,\mu\text{s}$ duration regime remains, as yet, a poorly explored domain of observations due to mainly the hardware limitations of space-based gamma-ray detectors and the tiny exposures of ground-based gamma-ray detectors. It is thus possible that there are classes of objects emitting such signals that have escaped detection. In other words, there is an opportunity for discovery in this observational space.

\subsection{Wavefront Technique}
\label{ss:wavefronttechnique}

While searches for sub-$10\,\mu\text{s}$ gamma-ray bursts are still very limited, there exists a technique suitable for detecting ultrashort bursts devised by Porter and Weekes for a pair of non-imaging Cherenkov detectors \cite{1978MNRAS.183..205P}, developed further for imaging telescopes \cite{Krennrich:1999tu,LeBohec:2002hq}, and realized in SGARFACE \cite{Schroedter:2008tp,2003ICRC....5.2971L} and briefly in SGARFACE+VERITAS \cite{Schroedter:2009iw,2009arXiv0908.0182S,2005APh....23..235L,2001AIPC..558..574K}. These studies were motivated by the prospects of detecting the possible gamma-ray counterparts to $\text{ns}-\mu\text{s}$ radio pulses from the Crab pulsar \cite{Schroedter:2009iw, Hankins:2007fa, Lyutikov:2007xw, Lyutikov:2016ueh, Bilous:2009jr} and explosive evaporation of primordial black holes (PBHs) as predicted by the (outdated) Hagedorn model \cite{Schroedter:2008tp}.\footnote{Within the Standard Model the explosive evaporation of a PBH is expected to last for hundreds of seconds \cite{Picker:2023ybp,Boluna:2023jlo}. The Hagedorn model was proposed before the Standard Model was experimentally confirmed. It predicts that, as the Hawking temperature of a black hole approaches the QCD scale of $\sim 200\MeV$, the number of states available for Hawking evaporation would increase exponentially, and this would accelerate the evaporation process, resulting in a burst of gamma-ray photons with a much shorter duration of hundreds of nanoseconds \cite{Halzen:1991uw}.} To keep the distinction clear, we refer to this detection scheme as the gamma-ray \textit{wavefront technique}. We briefly highlight important aspects of this technique here and detail it further in Appendix.~\ref{appendix:wavefronttechnique}.\footnote{Note that the \textit{gamma-ray} wavefront technique is to be distinguished from the so-called \textit{Cherenkov} wavefront sampling technique, an outdated technique to detect \textit{individual} gamma-rays that was implemented in detectors such as ASGAT, Themistocle, CELESTE, and STACEE \cite{deNaurois:2015oda}.}

Gamma-ray photons from conventional sources arrive at Earth well separated spatially and are observed one at a time at IACTs through their Cherenkov images \cite{2023hxga.book..144E,Prandini:2022wcb,Bose:2021oiq}. By contrast, an ultrashort gamma-ray burst we have in mind would arrive with a high fluence in the form of a thin, planar wavefront sweeping through space, which creates a large number of overlapping showers when it enters the Earth's atmosphere. The rough condition for overlapping showers is that the primary gamma-ray fluence $\mathcal{F}_*$ is much greater than $10^{-5}\,\text{ ph}/\text{ m}^{-2}$, which corresponds to the inverse of the typical ground area $A_{\rm pool}\sim 10^{5}\text{ m}^2$ of a gamma-ray induced shower. Since the Cherenkov light from these showers are fully mixed up in this case, the primary gamma-rays are not detected individually but collectively \cite{1978MNRAS.183..205P,Schroedter:2008tp,Schroedter:2009iw,2005APh....23..235L}. This is the key difference between the wavefront technique and the standard detection schemes of IACTs.

Extensive Monte Carlo simulations show that the Cherenkov images from the superimposed showers initiated by a gamma-ray wavefront are distinct, both morphologically and temporally, from that which result from a single gamma-ray \cite{Krennrich:1999tu,LeBohec:2002hq,Schroedter:2008tp}. Single gamma-ray-induced images are elliptical, exhibit parallactic variations among telescopes due to the different distances of the telescopes to the shower core, and can only be detected within the $\sim 150\text{ m}$ pool radius of the shower. On the other hand, gamma-ray-wavefront-induced showers do not have a well-defined core and instead extend more or less uniformly on the ground. They create images that are much more circular and would appear nearly identically in many telescopes spread over vast distances. Further, while the typical Cherenkov flash of a single gamma-ray induced shower is a few nanoseconds long, the time profile of the Cherenkov light of gamma-ray wavefront showers may last for $\gtrsim 10\text{ ns}$ \cite{Krennrich:1999tu,LeBohec:2002hq,Schroedter:2008tp}, reflecting the intrinsic duration of the gamma-ray burst. The wavefront technique utilizes these distinct properties of the Cherenkov images that result from gamma-ray wavefronts entering the atmosphere to achieve essentially background-free detection of ultrashort gamma-ray bursts.

A major challenge in detecting single or wavefront gamma-ray induced shower signals with IACTs is achieving sufficient rejection of background events, mainly due cosmic-ray-induced showers and the light of the night sky. A signal-background separation can, in principle, be made based on the recorded camera images and their time development. In practice, current-generation IACTs reject backgrounds in two steps:
\begin{enumerate}
    \item \textit{Online Triggering}\\
    In order not to overwhelm the data acquisition system, the majority of the backgrounds are already rejected online (in real time, before the data are read out) by applying a three-level trigger system, usually dubbed Level 1, 2, and 3. The single-pixel (Level 1) trigger checks if a threshold number of counts is registered in each camera pixel within a certain time window, enabling selection based on the temporal profile of the event. Next, the multi-pixel (Level 2) trigger checks for coincident Level 1 triggers in multiple adjacent pixels within a specified time window, enabling selection based on the angular size of the event. Finally, the multi-telescope (Level 3) trigger checks for coincident Level 2 triggers in at least two telescopes within a time window, providing selection based on the spatial size of the event on the ground.     
    \item \textit{Offline Reconstruction}\\ 
    Events that pass the Level 3 trigger are read out by the data acquisition system and stored in the memory after some delay, typically $\mathcal{O}(10\,\mu\text{s})$. The latter contributes to the dead time of the detector.\footnote{The dead time of IACTs is not an important limiting factor for detecting ultrashort gamma-ray bursts with the wavefront technique because it takes only one image, i.e. one event, to recognize a burst.} Further background rejections and reconstruction of the primary gamma-rays are done offline by comparing the recorded images with expected signal images as found in Monte Carlo simulations.
\end{enumerate}
For more details on the trigger and data acquisition systems of IACTs, see e.g. \cite{Paoletti2004,Funk:2004ie,Weinstein:2007ss}.

The standard Level 1 trigger employed in existing and planned IACTs typically integrates over a time window of $\mathcal{O}(10\text{ ns})$. Such a short trigger window is optimized for selecting single gamma-ray shower events whose durations are $\mathcal{O}(10\text{ ns})$. To maximize the sensitivity to ultrashort gamma-ray burst events that last $\gtrsim 10\text{ ns}$, different Level 1 integration times are warranted. Ideally, the integration time should match the duration of the signal \cite{2001ICRC....7.2756L}.\footnote{With shorter integration time, the sensitivity degrades because only a fraction of the signal can fit into an integration window; with longer integration time, the sensitivity suffers from both signal dilution and background exaggeration.} Since the burst duration is \textit{a priori} unknown, it is useful to have a trigger system that is sensitive to various burst durations at once. This is the idea behind the multi-time scale discriminator (MTSD) of the SGARFACE Level 1 trigger, which integrates the signal on six integration windows: 60 ns, 180 ns, 540 ns, 1620 ns, 4860 ns, and 14580 ns \cite{Schroedter:2008tp,Schroedter:2009iw}. Employing an MTSD-like trigger not only achieves near-optimal trigger sensitivity to a wide range of burst durations longer than $10\text{ ns}$ but also enables better rejection of frequent 5-30 ns Cherenkov flashes from cosmic-ray showers.

Even if the temporal profile of the gamma-ray burst is a delta function, random air shower process inevitably introduces $\mathcal{O}(10\text{ ns})$ smearing of the arrival times of Cherenkov photons at the telescope \cite{Krennrich:1999tu,LeBohec:2002hq,Schroedter:2008tp}.\footnote{The time spread in the arrival times of the Cherenkov photons remains small, 
despite the $\mathcal{O}(10\text{ km})$ uncertainty in the heights at which a shower could start, because the showers develop at nearly the speed of light and are highly beamed. Given that the typical angular spread of Cherenkov light from a wavefront event is $\theta_{\rm Ch}\sim 1^\circ$, we can estimate the time spread of the Cherenkov light to be $10\text{ km}\left(1-\cos\theta_{\rm Ch}\right)\sim 10\text{ ns}$, which explains the time spread found in the Monte Carlo simulations of \cite{Krennrich:1999tu,LeBohec:2002hq,Schroedter:2008tp}. Note also that the geometry of a $\sim 10\text{ m}$ aperture telescope may also introduce a time spread of similar order due to the different path lengths of rays depending on where they hit the mirror.} It follows that primary gamma-ray bursts with any durations shorter than $\sim 10\text{ ns}$ would produce virtually the same Cherenkov photon signals, independent of the burst duration. Detecting such sub-10 ns bursts might be relatively challenging compared to detecting longer-duration ones because their online selection necessitates short, $\mathcal{O}(10\text{ ns})$, integration times at Level 1 which result in poor temporal rejection of cosmic ray backgrounds at trigger level based on arrival times of photons. While further simulation studies are required, it may still be possible to identify $\lesssim 10\text{ ns}$ gamma-ray burst signals among the cosmic-ray shower background without relying on the temporal discrimination, i.e. based on mainly Level 2 and 3 triggers and offline analysis of the Cherenkov images. This is more likely to work if a large number of telescopes are used for stereoscopic rejection at Level 3.\footnote{The past decade has seen the advent of deep learning techniques to deal with data in the form of sequences of images. These techniques have been demonstrated to improve event reconstruction \cite{2022ASPC..532..191N,2021arXiv211201828M} and background suppression \cite{Spencer:2021ypz,2019ICRC...36..752N,Parsons:2019myj,2019JPhCS1181a2048P,2017APh....89....1K} in the analysis of simulated IACT data. Similar techniques may be applied to help achieve background-free search for gamma-ray wavefront events.}

\subsection{Expanding the Search for Rare Ultrashort Gamma-Ray Bursts}
Although the wavefront technique was developed more than two decades ago, gamma-ray bursts at sub-$10\,\mu\text{s}$ timescales have not been searched for with good sensitivity and exposure. SGARFACE has carried out a 1502 hr search for ultrashort bursts but relying only on Level 1 and Level 2 triggers in a single Whipple telescope \cite{Schroedter:2008tp}. The lack of stereoscopic (Level 3) rejection in this search leads to poor rejection of cosmic ray backgrounds. SGARFACE and VERITAS have carried out a stereoscopic search with multiple telescopes, achieving excellent Level 3 background rejection, but only for 6.3 hr \cite{Schroedter:2009iw}. On top of that, these telescopes have poor FoVs.

We propose three (economical) efforts to extend the search for ultrashort gamma-ray bursts in the near future:
(1) by analyzing the archival data of SGARFACE, VERITAS, and other IACTs
(2) by installing piggyback trigger systems in existing and planned IACTs (3) by using PANOSETI, an upcoming all-sky, all-time survey for fast optical transients.

\subsubsection{Searches in the archival data of existing IACTs}

If the signals of our interest pass the trigger criteria of existing IACTs, such signals might already be present in their archival data. In fact, SGARFACE, which is optimized for $\sim 60\text{ ns}-15\,\mu\text{s}$ duration bursts, has found, after a 1502-hour search, 14 events that passed its Level 2 trigger and cannot be ruled out as sub-$10\,\mu\text{s}$ gamma-ray wavefront events \cite{Schroedter:2008tp}. While stereoscopic observation with one or more additional telescopes would most probably have rejected many of these events, it is possible that our signals have been hiding among these events. 

The standard trigger systems employed in current-generation IACTs, e.g., VERITAS \cite{Adams:2021hzq}, have $\mathcal{O}(10\text{ ns})$ Level 1 integration windows. Such integration windows should at least be able to catch bursts with durations not much longer than $10\text{ ns}$. Although some of our signals seem to pass the standard triggers of IACTs, their subsequent standard data analysis chains \cite{Daniel:2007wd} do not search for Cherenkov images associated with gamma-ray wavefront events. Archival data searches for gamma-ray bursts have been done in the past, but these searches focused on conventional, longer-lasting sources whose gamma-rays are detected individually and $\gtrsim 10^3\text{ s}$ integration time windows are needed to achieve sufficient signal-to-noise ratios \cite{Krennrich:1999tu,LeBohec:2002hq,Schroedter:2008tp,Tesic:2012kx,Skole2016Search,Archambault2016Search,Archambault:2017asc,Doro:2021dzh}. It is thus worthwhile to re-analyze the archival data of IACTs to search specifically for wavefront events.

\subsubsection{Piggyback trigger systems}

The online trigger systems in existing and planned IACTs are optimized for conventional sources. These standard trigger systems may not be optimal for selecting putative ultrashort gamma-ray bursts. It is possible that our signals are vetoed at the trigger level, and thus not recorded. In that case, a specific trigger system that better selects these events should be installed in IACTs. Building new telescopes that are entirely dedicated to exploratory ultrashort gamma-ray burst searches may be expensive and not cost effective. We instead propose installing an MTSD-like trigger that integrates over multiple time windows longer than $10\text{ ns}$ in existing and planned IACTs in order to maximize their trigger sensitivities for a wide range of burst durations.

As demonstrated by the SGARFACE collaboration, existing IACTs can be made more sensitive to ultrashort gamma-ray bursts by simply having a new trigger system installed in it. The SGARFACE collaboration installed a new trigger system for the Whipple telescope in order to improve its sensitivity to longer-duration bursts, without interfering with the standard operation of the telescope \cite{Schroedter:2008tp,Schroedter:2009iw}. This can be done by splitting the PMT signals before they reach the standard trigger system and running the copied signal through the new trigger system. Originally, the Whipple telescope employed a trigger system that integrates over an $\mathcal{O}(10)\text{ ns}$ time window \cite{Doro:2021dzh}, which would capture only a small fraction of a $\gg 10\text{ ns}$ burst signal. The SGARFACE trigger system, on the other hand, integrates over multiple time windows, the shortest of which is 60 ns (three consecutive intervals of 20 ns), thus greatly enhancing the Whipple telescope's ability to probe bursts with longer durations.

In a way similar to how the SGARFACE system operates on the data collected by the Whipple telescope, a new trigger system with improved sensitivity to ultrashort bursts can piggyback on existing telescopes without affecting the standard operation of the telescope. For example, the piggyback trigger can be installed on VERITAS, MAGIC, HESS, and also Cherenkov Telescope Array (CTA) in the near future. The $\sim 3-5^\circ$ FoV diameter of an IACT is usually significantly bigger than both its $\sim 0.1^\circ$ angular resolution and the angular extent of the object being observed, which means most of the pixels of the IACT would be empty of photoelectrons most of the time \cite{Cassanyes:2015wpr}. A Cherenkov flash from an ultrashort gamma-ray burst could appear in the FoV of the telescope even when it is performing a scheduled observation of another object. Hence, the full $\sim 1000\text{ hr}/\text{yr}$ data-taking time of an IACT can, in principle, be used to simultaneously search for ultrashort bursts, if a suitable trigger system is running. In this way, we can carry out our proposed search in parallel with the standard analysis routine of the telescope.

\subsubsection{All-sky all-time searches with PANOSETI}
Another challenge with detecting gamma-ray bursts of the type we have in mind has to do with their possibly infrequent---perhaps once a year---rate of occurrence. The signal would be absent most of the time and when a burst occurs, it would appear for a short duration at a random point on the sky. An effective search for such rare and very fast gamma-ray transients requires detectors with not only short time resolutions, but also with large instantaneous field of coverage and operating continuously for a long observation time. 

The Pulsed All-sky Near-Infrared Optical SETI (PANOSETI) seems to have all these qualities and is currently in its final design stage \cite{2022SPIE12184E..8BM}. Unlike most optical sky surveys (Pan STARRS, Zwicky Transient Factory, Vera C. Rubin Observatory) which have integration times of several minutes or longer \cite{2019BAAS...51g.264W}, PANOSETI will probe the optical and near-infrared band with 10 nanosecond resolution. While PANOSETI is developed to search for fast-transient optical SETI targets, they can pull double duty as IACTs \cite{Korzoun:2023jgb}. Two prototype PANOSETI telescopes were tested in conjunction with VERITAS for about four nights. It was demonstrated that they could detect showers induced by 15-50 TeV gamma-ray photons from the Crab nebula. PANOSETI's significantly larger FoV and nearly $100\%$ duty cycle sets it apart from current gamma-ray detectors in its ability to survey the sky for ultrashort gamma-ray transients that occur rarely.

\subsection{Projected Sensitivities}
\label{ss:projectedsensitivity}
\begin{figure*}
    \centering
    \includegraphics[width=0.49\linewidth]{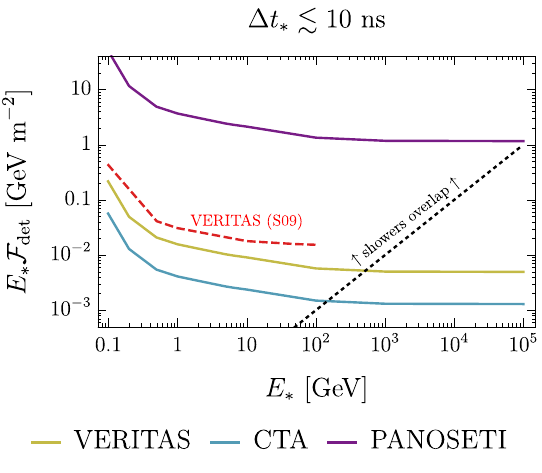}
    \includegraphics[width=0.49\linewidth]{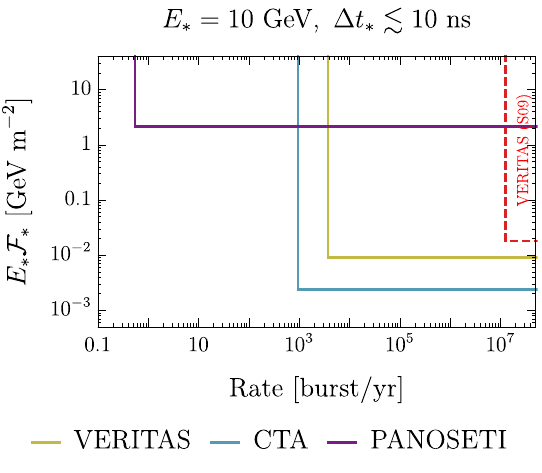}\\
   \includegraphics[width=0.49\linewidth]{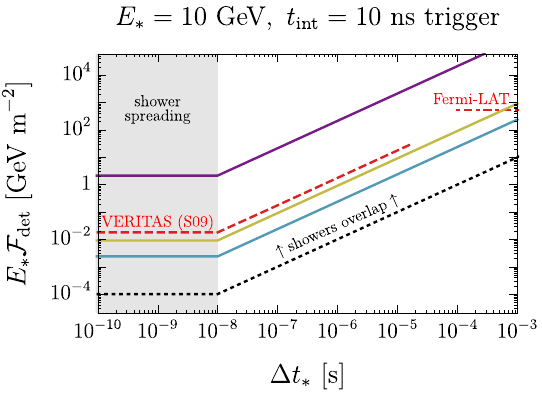} \includegraphics[width=0.49\linewidth]{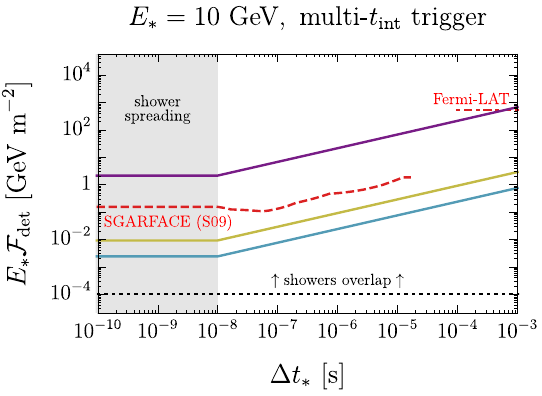}
    \caption{Projected sensitivities of VERITAS, CTA, and PANOSETI to ultrashort gamma-ray bursts, assuming a background-free search can be achieved with the wavefront technique (see Appendix.~\ref{appendix:wavefronttechnique}). The spectra of the bursts are assumed monochromatic at an energy $E_*$ and the sensitivities are shown in terms of the energy fluence of the burst $E_*\mathcal{F}_*$. Shown for comparison in dashed red lines labeled VERITAS (S09) and SGARFACE (S09), respectively, are the sensitivities of VERITAS and SGARFACE (separately) as reported in \cite{Schroedter:2009iw}. The $\mathcal{O}(1)$ difference between the VERITAS (S09) line and our projected VERITAS sensitivity is due to the PMT upgrade of VERITAS reported in \cite{2011ICRC....9...14K,2011arXiv1110.4702N}. Also shown for comparison in dot-dashed lines are the sensitivity of Fermi-LAT, whose detection threshold is assumed here to correspond to consecutive event triggers by at least five photons crossing within its effective collection area of $0.95 \text{ m}^2$ over a time period longer than $100\,\mu\text{s}$ (roughly four times the dead time of Fermi-LAT) \cite{Atwood_2009}. The dotted lines correspond to the gamma-ray fluence $\mathcal{F}_*$ satisfying $\mathcal{F_*} t_{\rm int}/\text{max}\left[\Delta t_*, 10\text{ ns}\right]=10^{-5}\text{ ph }\text{ m}^{-2}$ which mark the rough boundaries above which the Cherenkov image captured within an integration time $t_{\rm int}$ receives contributions from numerous showers instead of just a single shower; below the dotted lines, the wavefront technique does not apply, see Section.~\ref{ss:wavefronttechnique} and Appendix.~\ref{appendix:wavefronttechnique}. \textit{Top Left: } photon energy $E_*$ dependence of the energy fluence sensitivities $E_*\mathcal{F}_{\rm det}$. \textit{Top Right: } the burst occurrence rates at which the energy fluence sensitivities shown in the other panels apply. While for concreteness the threshold values of $E_*\mathcal{F}_*$ are shown for photon energy $E_*=10\GeV$ and burst duration $\Delta t_*\lesssim 10\text{ ns}$, the threshold burst rates depend on neither $E_*$ nor $\Delta t_*$. \textit{Bottom Left and Right: }burst duration $\Delta t_*$ dependence of the energy fluence sensitivities $E_*\mathcal{F}_{\rm det}$ for a trigger system with a single integration time of $t_{\rm int}=10\text{ ns}$ (similar to that employed by VERITAS \cite{Schroedter:2009iw}) and for a trigger system with multiple integration times $t_{\rm int}$ ranging from $10^{-8}\text{ s}$ to $10^{-3}\text{ s}$ [similar in spirit to the multi-time-scale discriminator (MTSD) employed by SGARFACE \cite{Schroedter:2009iw}], respectively. Because the randomness of air showers introduces $\mathcal{O}(10\text{ ns})$ time spreads in the time developments of the Cherenkov images in these detectors, bursts with durations $\Delta t_*\lesssim 10\text{ ns}$ result in virtually the same Cherenkov time profiles, hence the flattening of the $E_*\mathcal{F}_{\rm det}$ at $\Delta t_*\lesssim 10\text{ ns}$.}
    \label{fig:fluencesensitivity}
\end{figure*}

The search for coincident gamma-ray wavefront events in both VERITAS and SGARFACE done in \cite{Schroedter:2009iw} is perhaps the only experiment performed thus far that has the capability of discovering $\lesssim 10\,\mu\text{s}$ gamma-ray bursts. They found no coincident gamma-ray wavefront events during a 6.3 hour run \cite{Schroedter:2009iw}, and thereby placed a limit on gamma-ray bursts with rates greater than $10^7\text{ burst/year}$.

To illustrate our near-future ability to extend the search for ultrashort gamma-ray bursts, we consider three representative detectors: VERITAS, CTA, and PANOSETI. VERITAS is an existing current-generation IACT which has been running since 2007 \cite{2023hxga.book..144E,Prandini:2022wcb,Bose:2021oiq} and is the only experiment apart from Whipple (with SGARFACE trigger), its predecessor, that has been used specifically to search for sub-$10\,\mu\text{s}$ gamma-ray bursts \cite{Schroedter:2009iw}, albeit only for one night. We estimate the fluence sensitivity of VERITAS to wavefront events (for burst durations $\Delta t_*\lesssim 10\text{ ns}$) in Appendix~\ref{appendix:wavefronttechnique}. CTA, a next-generation IACT that is currently being built, is expected to reach about an order of magnitude of improvement in energy fluence $E_*\mathcal{F}_{\rm det}(E_*)$ as well as slightly better FoV and duty cycle compared to current-generation IACTs. The significant increase in the fluence sensitivity will enable us to better probe fainter bursts. PANOSETI uses a large number of low-cost telescopes with relatively small collecting area to cover a larger FoV. Essentially, it trades sensitivity for better exposure, thus making it suitable for probing relatively strong bursts that occur infrequently. We estimate the energy fluence sensitivities of CTA and PANOSETI based on simple adaptation of the procedure for VERITAS, which we also detail in  Appendix~\ref{appendix:wavefronttechnique}. The results are shown in Fig.~\ref{fig:fluencesensitivity}.

Assuming the bursts occur at a constant and uniform rate $\Gamma_{\rm burst}$ over the whole sky, the probability of catching a burst within the field of view $\Omega_{\rm FoV}$ and data-taking time of a detector $T_{\rm obs}$ is $\Gamma_{\rm burst}T_{\rm obs}\left(\Omega_{\rm FoV}/4\pi\right)$. Requiring this probability be greater than unity sets the minimum burst rate $\Gamma_{\rm burst}$ that the detector is sensitive to. The rate sensitivities of the three detectors we consider as per this criterion are displayed in Fig.~\ref{fig:fluencesensitivity}. The assumed field of view solid angles $\Omega_{\rm FoV}$ of VERITAS, CTA, and PANOSETI are $2.9\times 10^{-3}\text{ sr}$, $2\times 4.4\times 10^{-3}\text{ sr}$,\footnote{The factor of 2 accounts for the northern and southern arrays \cite{Hofmann:2023fsn}.} and $2.27\text{ sr}$, respectively. Moreover, we assume that the detectors under consideration run for approximately ten years (the typical lifetime of an experimental project, not accounting for duty cycle). This amounts to $T_{\rm obs}=10\times 1000\text{ hr}$ for VERITAS \cite{2014NIMPA.766...61R} and $T_{\rm obs}=10\times 1300\text{ hr}$ for CTA \cite{Hofmann:2023fsn} given the $\sim \mathcal{O}(10)\%$ duty cycle of IACTs, and $T_{\rm obs}=10\times 8766\text{ hr}$ for PANOSETI which we assume to have $100\%$ duty cycle. Further details of these detectors can be found in Appendix.~\ref{appendix:wavefronttechnique}.

For longer burst durations, $\Delta t_*\gtrsim 10\text{ ns}$, the energy fluence threshold for a detection $E_*\mathcal{F}_{\rm det}$ varies with $\Delta t_*$ in a way that depends on both the trigger algorithm and the background rate. It scales as $E_*\mathcal{F}_{\rm det}\propto \Delta t_*$ for VERITAS, whereas it scales as $E_*\mathcal{F}_{\rm det}\propto \sqrt{\Delta t_*}$ for SGARFACE owing to its MTSD \cite{Schroedter:2009iw,Schroedter:2008tp}. In Figure.~\ref{fig:fluencesensitivity}, we plot both the $E_*\mathcal{F}_{\rm det}\propto \Delta t_*$ and $E_*\mathcal{F}_{\rm det}\propto \sqrt{\Delta t_*}$ extrapolations of the $E_*\mathcal{F}_{\rm det}$ obtained in Appendix~\ref{appendix:wavefronttechnique} for $\Delta t_*\lesssim 10\text{ ns}$ to longer burst durations.

The linear in $\Delta t_*$ scaling of the energy fluence sensitivity of VERITAS can be understood as follows. VERITAS employs a trigger system with a \textit{single} integration time window of $t_{\rm int}=7\text{ ns}$, which means only a fraction $t_{\rm int}/\Delta t_*$ of the fluence of a $\Delta t_*\gtrsim t_{\rm int}$ gamma-ray burst signal would lie inside an integration interval $t_{\rm int}$. In order to meet the same photoelectron threshold of the PMTs, the gamma-ray fluence would need to be larger by a factor $\Delta t_*/t_{\rm int}$, thus explaining the $E_*\mathcal{F}_{\rm det}\propto \Delta t_*$ scaling. Catching the entirety of the burst signal is possible if the integration time $t_{\rm int}$ is longer than the burst duration $\Delta t_*$. However, it is disadvantageous to make $t_{\rm int}$ arbitrarily long as doing so would increase the background count, which typically scales as $\sqrt{t_{\rm int}}$. The optimal trigger threshold is achieved when the integration time matches the burst duration, $t_{\rm int}\approx \Delta t_*$. While we do not \textit{a priori} know the $\Delta t_*$, such an optimal sensitivity can be achieved simultaneously for a broad range of burst durations $\Delta t_*$ with the use of a trigger system that integrates over \textit{multiple} integration times $t_{\rm int}$, similar to the MTSD of SGARFACE. As explained in the preceding two subsections, such a trigger system can work in parallel without interfering with the standard operation of the telescopes. If an MTSD-like trigger is installed in each of the three detectors under consideration--VERITAS, CTA, and PANOSETI--we expect their energy fluence sensitivities to scale as $E_*\mathcal{F}_{\rm det}\propto \sqrt{t_{\rm int}}\approx \sqrt{\Delta t_*}$, as is the case for SGARFACE.\footnote{ Note that the shortest integration time of the MTSD of SGARFACE is $t_{\rm int}=3\times 20\text{ ns}$, which explains why the $\propto \sqrt{\Delta t_*}$ scaling of its energy fluence sensitivity pivots at $\Delta t_*\approx 60\text{ ns}$. In our extrapolation of the sensitivities of VERITAS, CTA, and PANOSETI shown in Fig.~\ref{fig:fluencesensitivity} we assume that the MTSD-like triggers for these telescopes have a minimum integration time of $10\text{ ns}$.}

We reiterate that the wavefront technique on which we base our analysis relies on detecting a smooth, superimposed Cherenkov image from a \textit{large number} of contributing showers that are initiated by the same gamma-ray wavefront. The applicability of this technique thus requires a sufficiently high gamma-ray fluence: $\mathcal{F}\gg A_{\rm pool}^{-1}\sim 10^{-5}\text{ ph }\text{ m}^{-2}$; see Section~\ref{ss:wavefronttechnique} and Appendix~\ref{appendix:wavefronttechnique}. Now, the relevant fluence $\mathcal{F}$ to satisfy this condition is the fluence that arrives on the ground within the integration time $t_{\rm int}$ of the detector, which is not necessarily the same as the full fluence of the burst $\mathcal{F}_*$. For $t_{\rm int}\lesssim \text{max}\left[\Delta t_*,10\text{ ns}\right]$, where the $10\text{ ns}$ accounts for shower spreading, we have $\mathcal{F}=\mathcal{F}_*$, and so the overlapping-shower condition amounts to $\mathcal{F}_*\gg 10^{-5}\text{ ph }\text{ m}^{-2}$. However, for $t_{\rm int}\gtrsim \text{max}\left[\Delta t_*,10\text{ ns}\right]$, only a fraction $t_{\rm int}/\text{max}\left[\Delta t_*,10\text{ ns}\right]$ of the full fluence $\mathcal{F}_*$ contributes to creating a Cherenkov image that is captured within an integration time $t_{\rm int}$, which translates to a stricter condition $\mathcal{F}_*t_{\rm int}/\text{max}\left[\Delta t_*,10\text{ ns}\right]\gg 10^{-5}\text{ ph }\text{ m}^{-2}$. The latter implies a lower bound on $\mathcal{F_*}$ for the wavefront technique to apply, that for $\Delta t_*\gtrsim 10\text{ ns}$ is linear in $\Delta t_*$ for a single integration time $t_{\rm}=10\text{ ns}$ trigger and independent of $\Delta t_*$ for a multi-$t_{\rm int}$ (MTSD-like) trigger; see Fig.~\ref{fig:fluencesensitivity}.

\section{Electromagnetic Signals of Dark Blob Collisions}
\label{section:EMsignals}

An intense gamma ray burst with duration $\lesssim 10\,\mu\text{s}$ implies an emitting blob of size $\lesssim 3\text{ km}$ (whose light travel time is $\lesssim 10\,\mu\text{s}$) with very high energy density, possibly higher than that of any known Standard Model object, that glows in gamma-ray for $\lesssim 10\,\mu\text{s}$ when the time is ripe. It is plausible that the dark matter comes in the form of such ultra-dense blobs that release an ultrashort burst of gamma-rays when they collide.

Nearly all known mechanisms for dark blob formation, which we summarize in the Section~\ref{ss:formation mechanism}, require the existence of some particles that mediate interactions between the fundamental dark matter particles constituting a blob. When a pair of dark blobs collide in the present epoch, a large number of these mediators may be radiated as an energy loss channel. If the mediators are coupled to the Standard Model in some way, the emitted mediators may subsequently decay with some branching ratio into photons. Even if this branching ratio to photon is small, non-photon decay products may thermalize into a fireball which then releases most of its energy in photons when it becomes optically thin (see Appendix.~\ref{appendix:fireball}). It is thus reasonable to expect a burst of photons to be produced in a blob collision. We present a model that does exactly that in Section.~\ref{ss:modelsummary} and Appendix.~\ref{appendix:modeldetails}. In this section, we discuss in a model agnostic way the prospect for discovering gamma-ray transients from dark blob collisions.

\subsection{Assumptions and Parameterization}

We assume the entire dark matter mass density comes in the form of identical dark blobs with mass $M$. An upper bound on the blob mass is given by microlensing constraints \cite{Smyth:2019whb,Jacobs:2014yca, Croon:2020wpr}, 
\begin{align}
    M\lesssim 10^{23}\text{ g}    \label{eq:microlensing}
\end{align}
For simplicity, we  write the velocity-averaged blob collision cross section in the Milky Way as
\begin{align}
    \left<\sigma v\right>_{\rm col}=v_{\rm DM}R^2 \label{eq:geometricalcrosssection}
\end{align}
where $v_{\rm DM}=10^{-3}$ is the typical virial velocity in the Milky Way and $R$ is the effective radius of the blob as defined by the above relation.\footnote{What we refer to as the blob radius $R$ merely parametrizes the square root of the blob collision cross section. If the blob collision cross section were Sommerfeld enhanced, for example, the actual blob radius could be much smaller.} The blob radius $R$ is bounded from above and below by requiring, respectively, consistency with Bullet Cluster observations ($R^2/M\lesssim 1\text{ cm}^2/\text{g}$)\footnote{Requiring the probability that over the age of the universe each blob collides less than once (to ensure, in some models, that the blobs are not depleted) yields a parametrically similar constraint to the Bullet Cluster one.} \cite{Jacobs:2014yca} and that the blobs not be black holes ($R> 2GM$), where $G$ is Newton's gravitational constant. Given these constraints, a wide range of blob radii are still allowed
\begin{align}
     10^{-11}\text{ cm}\left(\frac{M}{10^{17}\text{ g}}\right)\lesssim  
     R\lesssim 10^8\text{ cm}\left(\frac{M}{10^{17}\text{ g}}\right)^{1/2} \label{eq:BCBH}
\end{align}
Variations in the outcomes of a collision due to different impact parameters are neglected in our analysis. We further assume that blob collisions occur only after Galaxy formation. The blobs may have formed before Cosmic Microwave Background (CMB) decoupling, however before structure formation we expect the blobs to move extremely slowly and are thus unlikely to collide \cite{Mack:2006gz}. Hence, there is no limits on blob parameters based on photon injection in the early universe and the resulting CMB spectral distortions \cite{Deng:2018cxb}.

To characterize a burst, we introduce three more parameters: 
\begin{itemize}
    \item  The fraction of $M$ released in gamma-ray $\epsilon_*$.\\
    We assume that each blob collision produces a gamma-ray burst with total energy $\epsilon_* M$. The $\epsilon_*$ parameter is, in principle, set by the coupling strength of the blob constituents to the mediator as well as other properties of the blob. Interesting values of $\epsilon_*$ include those corresponding to scenarios where the burst energy saturates the blob's mass ($\epsilon_*\sim 1$), the blob's virial kinetic energy ($\epsilon_*\sim v_{\rm DM}^2\sim 10^{-6}$), or the blob's binding energy ($\epsilon_*\sim v_{\rm esc}^2$, where $v_{\rm esc}$ is the escape velocity of a blob constituent from the confining potential of the blob). When showing our results in plots, we set $\epsilon_*=1$ and $\epsilon_*=10^{-6}$, with the understanding that the result for different values of $\epsilon_*$ can be obtained by simple rescalings. 
    \item Single gamma-ray energy $E_*$.\\
    We assume the bursts are monoenergetic. The individual photon's energy $E_*$ may be set by the mass, translational kinetic energy, thermal energy, or Fermi energy of a typical blob constituent. For simplicity, we will present our results only for $E_*$ in the range $100\MeV\leq E_*\lesssim 100\GeV$. This corresponds to the range of $E_*$ for which simulations of gamma-ray-wavefront-induced showers are currently available \cite{Krennrich:1999tu,LeBohec:2002hq,Schroedter:2008tp,2003ICRC....5.2971L,Schroedter:2009iw,2005APh....23..235L}. For $E_*$ above this range, we expect the wavefront technique to still be applicable, though the
    gamma-ray fluence $\mathcal{F}_*$ will be subject to a lower bound as per overlapping-shower requirement that is higher than the sensitivity of the detectors under consideration, complicating the analysis; see the last paragraph of Section.~\ref{ss:projectedsensitivity} and  Fig.~\ref{fig:fluencesensitivity}.
    \item The duration of the burst $\Delta t_*$.\\
    Depending on the nature of the collision, the burst duration $\Delta t_*$ may be set by the blob's light crossing time $R$, the blob interpenetration time $R/v_{\rm DM}$, the mediator's decay lifetime, or a model-dependent dynamical timescale (e.g. the blob collapse rate at the moment of photon production). As explained in Section.~\ref{section:detectiontechniques}, Cherenkov light signals as seen by ground-based gamma-ray detectors are independent of $\Delta t_*$ as long as $\Delta t_*\lesssim 10\text{ ns}$, which we assume to be the case in our blob parameter space plots.\footnote{Forward scatterings of gamma-rays with CMB photons and electrons in the interstellar medium may introduce additional time dispersions. The resulting time dispersions are in general extremely small \cite{Brevik:2020cky}. For Galactic sources we consider here, the time dispersions are many orders of magnitude less than $10\text{ ns}$, and so are negligible in our analysis. } 
\end{itemize}

An energy release in the form of Standard Model particles that is sufficiently concentrated both spatially and temporally may lead to the formation of a fireball, a thermalized, optically thick plasma that expands under its own radiation pressure. We consider this possibility in Appendix.~\ref{appendix:fireball} and find that the above parametrization also applies to the gamma-ray burst that is released from a fireball.

To further simplify our analysis, we confine ourselves to the regime where burst events originating in the Galaxy can be treated independently. Very crudely and regardless of the detector, this is the case if the $\Delta t_*$-thick shells of photons created in blob collisions do not have significant volume overlap in the Galaxy. In other words, the total volume occupied by burst shells created in the typical time it takes to escape the Milky Way, $\sim 10\text{ kpc}\sim 3\times 10^4\text{ yr}$, must be less than the volume of the Milky Way $\sim (10\text{ kpc})^3$, which boils down to 
\begin{align}
    \Gamma_{\rm col}^{\rm MW}\Delta t_*\lesssim 1
\end{align}
where $\Gamma_{\rm col}^{\rm MW}\sim \left(\rho_{\rm DM}/M\right)^2\times R^2v_{\rm DM}\times (10\text{ kpc})^3$ is the average collision rate of blobs in the Milky Way. This can be satisfied by choosing a sufficiently small $\Delta t_*$.

\subsection{Galactic Gamma-Ray Transients}
\label{ss:GalacticTransients}

\subsubsection{Visibility depth}

\begin{figure}
    \centering
    \includegraphics[width=\linewidth]{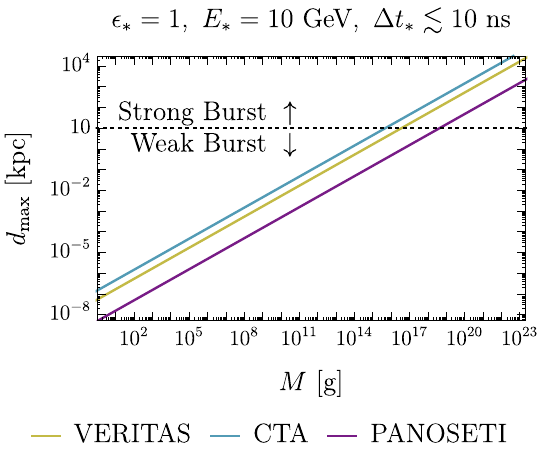}
    \caption{The characteristic visibility radii $d_{\rm max}$ of VERITAS, CTA, and PANOSETI to individual gamma-ray transients from dark blob collisions in the Milky Way as functions of the blob mass $M$. Each blob collision is assumed to produce a burst of gamma-rays with a duration $\Delta t_*$, a total energy $\epsilon_*M$, and a monochromatic energy per photon $E_*$, where $\Delta t_*\lesssim 10\text{ ns}$, $\epsilon_*=1$ and $E_*=10\GeV$ as labeled in the figure. The $d_{\rm max}$ is given explicitly by Eq.~\eqref{eq:dmax} and is computed based on the energy fluence sensitivities of the respective detectors as shown in Fig.~\ref{fig:fluencesensitivity}. The dotted line indicates the scale radius of the Milky Way halo and demarcates two distinct regimes for burst detection. Individual bursts originating from anywhere in the Galaxy are detectable in the strong burst regime ($d_{\rm max}\gtrsim 10\text{ kpc}$), whereas only bursts that occur sufficiently close can be detected in the weak burst regime ($d_{\rm max}\ll 10\text{ kpc}$).}
    \label{fig:dmax}
\end{figure}

The arriving energy fluence of the burst emitted in a blob collision at a distance $d$ from Earth is\footnote{Photons of a gamma-ray burst may hit other photons in the interstellar medium (mainly the CMB and infrared radiation by dust) and pair produce electron-positron pairs. Such gamma-ray attenuation effect is negligible in the cases we consider. Gamma-rays with energies $\lesssim 10\TeV$ originating from essentially anywhere in the Milky Way will reach Earth with near-unity survival probabilities \cite{Vernetto:2016alq}. At higher energies, gamma-rays may still survive with $\mathcal{O}(1)$ probabilities.}
\begin{align}
    E_*\mathcal{F}_*=\frac{\epsilon_* M}{4\pi d^2}
\end{align}
For a given detector, there is a minimum energy fluence above which a burst is detectable, $E_*\mathcal{F}_{\rm det}(E_*)$. Requiring $E_*\mathcal{F}_*>E_*\mathcal{F}_{\rm det}(E_*)$ translates to the maximum distance at which a blob collision is detectable
\begin{align}
    d_{\rm max}=\sqrt{\frac{\epsilon_* M}{4\pi E_*\mathcal{F}_{\rm det}(E_*)}} \label{eq:dmax}
\end{align}
Fig.~\ref{fig:dmax} shows the $d_{\rm max}$ as a function of blob mass $M$ for $\epsilon_*=1$, $E_*=10\GeV$, and the detectors considered in Section~\ref{ss:projectedsensitivity} whose energy fluence sensitivities $E_*\mathcal{F}_{\rm det}$ are displayed in Fig.~\ref{fig:fluencesensitivity}. As shown in Fig.~\ref{fig:dmax}, we can roughly demarcate two cases based on whether $d_{\rm max}$ is larger than 10 kpc (the length scale associated with the size of the Milky Way halo):
\begin{itemize}
    \item \textit{Strong burst}: If $d_{\rm max}\gtrsim 10\text{ kpc}$, then a blob collision occurring anywhere in the Milky Way is detectable as long as it happens within the search time and FoV of the detector. 
    \item \textit{Weak burst}: If $d_{\rm max}\ll 10\text{ kpc}$, then blob collisions are detectable if they happen sufficiently frequently in the Milky Way that at least one of them takes place at a distance less than $d_{\rm max}$ within the search time and FoV of the detector.
\end{itemize}

\subsubsection{Prospects for discovery}

The rate at which blob collisions occur is model dependent. It depends not only on the mass and radius of the blob, but also on whether the blobs form binaries and/or interact via long-range forces. Here, we simply assume that the blob collision cross section is given by Eq.~\eqref{eq:geometricalcrosssection}, where the role of the blob radius $R$ is to parametrize the collision rate. While we will call $R$ as defined by Eq.~\eqref{eq:geometricalcrosssection} the blob radius, it does not necessarily reflect the actual size of the blob. As such, the duration of the burst produced in a blob collision $\Delta t_*$ is \textit{a priori} independent of $R$.

The expected number of detectable blob collisions occurring within the field of view (FoV) of the detector $\Omega_{\rm FoV}$ during the observation time of the detector $T_{\rm obs}$ can be estimated as follows
\begin{align}
    N_{\rm event}= &J_{\rm col}\left(\frac{\rho_s}{M}\right)^2 R^2v_{\rm DM}T_{\rm obs}\Omega_{\rm FoV}\nonumber\\
    &\times \text{min}\left[d_{\rm max}^3,\left(10.6\text{ kpc}\right)^3\right] \label{eq:Nevent}
\end{align}   
with the criterion for detecting bursts from blob collisions being $N_{\rm event}>1$. Here
\begin{align}
    J_{\rm col}=&\frac{\int_{\rm FoV}d\Omega\, l^2dl\left(\rho_{\rm DM}^{\rm NFW}/\rho_s\right)^2}{\Omega_{\rm FoV}\times \text{min}\left[d_{\rm max}^3,\left(10.6\text{ kpc}\right)^3\right]} \label{eq:Jcol}
\end{align}
is an $\mathcal{O}(1)$ numerical factor to account for the spatial distribution of dark matter, $\Omega_{\rm FoV}$ is the FoV solid angle of the detector being used, $l$ is the line of sight distance, and we assume that the Galactic dark matter density $\rho_{\rm DM}^{\rm NFW}$ follows the NFW profile \cite{Navarro:1995iw}
\begin{align}
    \rho_{\rm DM}^{\rm NFW}(r)=\frac{\rho_s}{(r/r_s)(1+r/r_s)^2}
\end{align}
where $r$ is the Galactocentric radius, $\rho_s=9\times 10^{-3}M_\odot\text{pc}^{-3}$, and $r_s=16\text{ kpc}$ \cite{Navarro:1995iw,2015ApJS..216...29B}. In calculating $J_{\rm col}$, we take the distance from the Earth to the Galactic Center to be $r_\oplus=8\text{ kpc}$ and assume that the detector's FoV points perpendicularly to the Galactic plane. For $\theta_{\rm FoV}\lesssim 100^\circ$ (corresponding to an FoV solid angle of $\Omega_{\rm FoV}=2\pi\left[1-\cos\left(\theta_{\rm FoV}/2\right)\right]$), we find that 
typically $J_{\rm col}\approx 0.28$, unless $d_{\rm max}\sim 10.6\text{ kpc}$ where $J_{\rm col}$ dips down by up to a factor of $\sim 3$ compared to the typical value. Nonetheless, in obtaining our results we always compute the full integral in Eq.~\eqref{eq:Jcol}.

The condition $N_{\rm event}>1$ suggests the following figures of merit for burst detectability in the weak burst regime ($d_{\rm max}\ll 10\text{ kpc}$)
\begin{align}
\underbrace{\left[\epsilon_*^{3/2}M^{-1/2}R^2\right]}_{\text{blob parameters}}\times \underbrace{\left[E_*\mathcal{F}_{\rm det}(E_*)\right]^{-3/2}}_{\text{sensitivity}}\times \underbrace{\left[\Omega_{\rm FoV}T_{\rm obs}\right]}_{\text{exposure}} \label{eq:weakburstmerit}
\end{align}
and in the strong burst regime ($d_{\rm max}\gtrsim 10\text{ kpc}$)
\begin{align}
    \underbrace{\left[M^{-2}R^2\right]}_{\text{blob parameters}}\times \underbrace{\left[\Omega_{\rm FoV}T_{\rm obs}\right]}_{\text{exposure}} \label{eq:strongburstmerit}
\end{align}
These show that the fluence sensitivity $E_*\mathcal{F}_{\rm det}(E_*)$ is an important factor in the detectability of weak bursts. By contrast, the discovery potential for strong bursts is mainly limited by the exposure of the detector, namely, the product $\Omega_{\rm FoV}T_{\rm obs}$. In that case, we can afford to reduce the fluence sensitivity (increase $E_*\mathcal{F}_{\rm det}$) and decrease the visibility depth $d_{\rm max}$ if that is what it takes to improve exposure. This is reminiscent of an effective strategy for directly detecting dark blobs: where we are looking for rare but spectacular events, and the strategy is to maximize exposure at the expense of sensitivity, see e.g. \cite{Grabowska:2018lnd,Ebadi:2021cte}. Though we do not consider it here, single strong bursts from outside of the Milky Way may also be detectable in some cases.

\begin{figure*}[t!]
    \centering
    \includegraphics[width=0.49\linewidth]{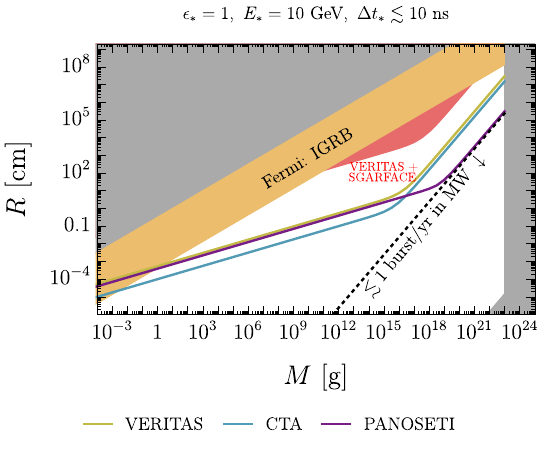}
    \includegraphics[width=0.49\linewidth]{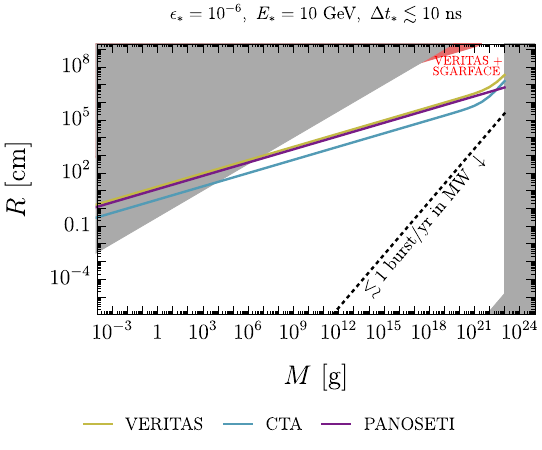}
    \caption{The radius $R$ vs mass $M$ parameter space of dark blobs. Shown in gray is the combined limits from Bullet Cluster (slanted upper boundary), microlensing (vertical boundary), and black hole avoidance (slanted lower boundary), cf.~Eqs.~\eqref{eq:microlensing} and \eqref{eq:BCBH}. The blobs are assumed to collide with a cross section $\left<\sigma v\right>_{\rm col}=10^{-3}R^2$ and produce upon a collision a burst of gamma-rays with a duration $\Delta t_*$, a total energy $\epsilon_*M$, and a monochromatic energy $E_*$ as labeled in each sub-figure. Along the dotted lines, blobs collide at the rate of once per year in the Milky Way. The colored solid lines are the projected sensitivities of VERITAS, CTA, and PANOSETI to the blob parameter space, obtained by requiring the expected number of detectable burst events $N_{\rm event}$ found in Eq.~\eqref{eq:Nevent} be greater than 1. The change in the slope of the sensitivity line from $R\propto M^{1/4}$ at relatively low $M$ to $R\propto M$ at relatively high $M$ corresponds to switching from weak burst regime (only sufficiently close bursts are detectable) to strong burst regime (bursts from any distance in the Galaxy are detectable); see Eqs.~\eqref{eq:weakburstmerit} and \eqref{eq:strongburstmerit}. The colored regions show the strongest existing indirect detection limits. For the case of $E_*=10\GeV$ shown here,
    these limits are from the isotropic diffuse gamma-ray background measurement by Fermi (Fermi: IGRB) \cite{Fermi-LAT:2014ryh} and 6.3-hour-long stereoscopic ultrashort gamma-ray burst search with VERITAS and SGARFACE (VERITAS+SGARFACE)  \cite{Schroedter:2009iw}.}
    \label{fig:moneyplot}
\end{figure*}

In Fig.~\ref{fig:moneyplot}, we show the projected sensitivities of VERITAS, CTA, and PANOSETI to $\epsilon_*=1$ and $\epsilon_*=10^{-6}$ bursts in terms of the blob parameters $(R,M)$. As we will discuss below, consistency with diffuse gamma-ray and cosmic ray observations already sets some limits on the blob parameter space. Nevertheless, a large parameter space that yields observable bursts in VERITAS, CTA, and PANOSETI still remains. For a given blob mass $M$, the constraints \eqref{eq:particletoblob} on $R$ derived from limits on particle dark matter annihilation cross sections scale as $\epsilon_*^{-1/2}$, whereas the figures of merit \eqref{eq:weakburstmerit} and \eqref{eq:strongburstmerit} imply that the threshold $R$ above which individual Galactic bursts can be detected scale as $\epsilon_*^{-3/4}$ in the weak burst regime and $\epsilon_*^{0}$ in the strong burst regime. It follows that as we decrease $\epsilon_*$ from $\epsilon_*=1$ to $\epsilon_*\ll 1$, the parameter space with observable Galactic bursts will shrink and expand in the weak and strong burst regime, respectively, until the $\epsilon_*$-independent upper bound on $R$ from Bullet Cluster dominates. Since $d_{\rm max}$ itself is proportional to $\epsilon_*^{1/2}$, the boundary between weak burst and strong burst, $d_{\rm max}\sim 10\text{ kpc}$, will also shift toward making the weak (strong) burst regime bigger (smaller) in the $(R,M)$ space.

\subsection{Diffuse Extragalactic Gamma-Ray Background}
\label{ss:IGRB}

\begin{figure}[t!]
    \centering
    \includegraphics[width=\linewidth]{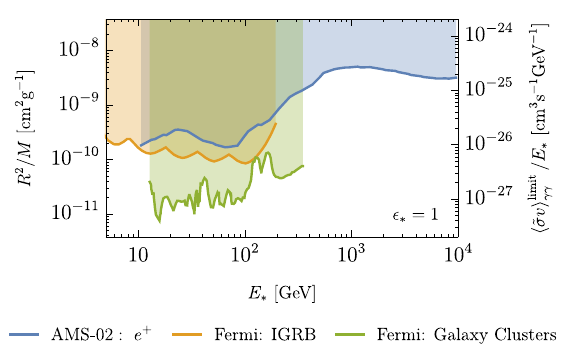}
    \caption{Existing indirect limits on the combination  $R^2/M$ (left vertical axis) of the blob radius $R$ and blob mass $M$ translated from limits on the annihilation cross section to two photons per unit mass $\left<\tilde{\sigma} v\right>_{\gamma\gamma}/\tilde{m}$ of particle dark matter with mass $\tilde{m}=E_*$, as per Eq.~\eqref{eq:particletoblob}. The limits shown are obtained from positron spectrum measurement by AMS-02 \cite{Elor:2015bho}, 
    isotropic diffuse gamma-ray background (IGRB) measurement by Fermi \cite{Fermi-LAT:2014ryh},
    and galaxy cluster observations by Fermi \cite{Anderson:2015dpc}. The blobs are assumed to collide with a cross section $\left<\sigma v\right>_{\rm col}=10^{-3}R^2$ and produce upon a collision a burst of gamma-rays with a total energy $\epsilon_*M$ with $\epsilon_*=1$ and a monochromatic energy per photon $E_*$.}
    \label{fig:ParticleLimitg}
\end{figure}

Some of the leading indirect detection constraints on annihilating \textit{particle} dark matter come from gamma-ray observations of the Galactic center and dwarf spheroidal galaxies \cite{Hooper:2018kfv,Slatyer:2017sev,Safdi:2022xkm,Gaskins:2016cha}. Nevertheless, in the case of dark blobs, due to the infrequent nature of their collisions (perhaps once per decade in the Milky Way, for example), these constraints either do not apply or need to be re-analyzed. Since we confine our analysis to the parameter space where dark blob collisions in the Milky Way produce non-overlapping bursts, gamma-ray observations that probe spatial volumes comparable to or smaller than the size of a Galaxy should expect temporally discrete, non-diffuse signals from blob collisions.

However, at galaxy cluster and cosmic scales, we expect collisions of blobs to likely lead to diffuse gamma-ray signals, in which case the usual constraints obtained in the case of particle dark matter annihilation should also apply for dark blobs. By equating the predicted photon flux in the two cases
\begin{align}
    \left(\frac{\rho_{\rm DM}}{M}\right)^2\left(v_{\rm DM}R^2\right)\left(\frac{\epsilon_*M}{E_*}\right)=\left[\left(\frac{\rho_{\rm DM}}{\tilde{m}}\right)^2\left<\tilde{\sigma} v\right>_{\gamma\gamma}\right]_{\tilde{m}=E_*}
\end{align}
where a tilde $\tilde{}$ denotes a particle dark matter quantity, we can recast limits on particle dark matter annihilation cross section to a pair of photons, $\left<\tilde{\sigma} v\right>_{\gamma\gamma}^{\rm limit}$, to the corresponding limits on the dark blob radius $R$ and mass $M$, 
\begin{align}
    \frac{R^2}{M}\lesssim \frac{\left<\tilde{\sigma} v\right>_{\gamma\gamma}^{\rm limit}}{\epsilon_*v_{\rm DM}E_*} \label{eq:particletoblob}
\end{align}
Fermi measurement of the isotropic diffuse gamma-ray background (IGRB) \cite{Fermi-LAT:2014ryh} and Fermi observation of the 16 highest J-factor galaxy clusters \cite{Anderson:2015dpc} have set $\left<\tilde{\sigma} v\right>_{\gamma\gamma}^{\rm limit}$ as function of the dark matter mass $\tilde{m}=E_*$, which we translate into limits on $R^2/M$ using Eq.~\eqref{eq:particletoblob}. These limits on $\left<\tilde{\sigma} v\right>_{\gamma\gamma}$ and the translated limits on $R^2/M$ are shown in Fig.~\ref{fig:ParticleLimitg} and have been incorporated into Fig.~\ref{fig:moneyplot}. 

While we assumed that the prompt products of a blob collision are monoenergetic photons, the final photon spectrum to be compared with data would generically have additional contributions from decaying pions, radiative processes such as final-state radiation and internal bremsstrahlung, and subsequent interactions with the surrounding medium (gas, starlight, CMB, cosmic rays, etc). We neglect the constraints on $R^2/M$ derived from those secondary spectra as line searches usually give the strongest constraints.

We note that gamma rays from blob collisions are clustered in both their arrival times (concentrated in intervals of $\Delta t_*$) and in their sky locations. This means the statistics of the arriving gamma-ray photons are inherently non-Poissonian, unlike in the case of particle dark matter, which predicts Poissonian statistics. This distinction can, in principle, be used to infer the presence of burst signals in the diffuse gamma-ray background. For instance, gamma-ray photons arriving from an angular patch in sub-$10\,\mu\text{s}$ bunches of a few photons more often than the Poisson expectation would indicate a non-Poissonian statistics. However, in practice, it might be difficult to detect the non-Poissonian nature of the photon statistics as it requires detecting more than one photon within the burst duration $\Delta t_*$. As discussed previously, Fermi-LAT, for instance, cannot register more than one photon within $\sim 10\,\mu\text{s}$. A good time resolution is thus essential for probing the non-Poissonian statistics.

\subsection{Cosmic Rays}

Generically, electrons and positrons should also be present among the final states that include monoenergetic photons. Positrons with energy $E\lesssim \TeV$ would lose a significant fraction of their energies in $\gtrsim 10^5\text{ yr}$ \cite{Joshi:2017ogv}. The current, steady-state positron population in the Galaxy probes the accumulation of $e^\pm$ injections from dark blob collisions over the long timescales set by the energy loss or diffusion timescale of positrons. In order to have a reasonable chance for detecting individual Galactic gamma-ray transients from blob collisions, we need the blob collision rate to be greater than about once per decade. Blob collisions at that rate would appear continuous over $\sim 10^{5}\text{ yr}$. Hence, an integration over such a long timescale washes out the discrete and burst-like nature of $e^\pm$ injections from blob collisions, the very features distinguishing them from that of particle dark matter. While it is difficult to extract telltale signature of blob collisions in cosmic ray data, constraints on the abundance of cosmic rays can be used to limit the blob parameter space. As before, such limits can be obtained by recasting cosmic-ray limits on the $\left<\sigma v\right>_{\gamma\gamma}$ of particle dark matter annihilation into two photons (which also produce cosmic rays automatically) into limits on $R^2/M$ using Eq.~\eqref{eq:particletoblob}. One such limit on $R^2/M$ can be obtained from the AMS-02 measurement of the cosmic-ray positron flux and the corresponding constraint on $\left<\sigma v\right>_{\gamma\gamma}$, as found in Ref.~\cite{Elor:2015bho}. This cosmic-ray limit is shown in Fig.~\ref{fig:ParticleLimitg} and has been incorporated into Fig.~\ref{fig:moneyplot}.

Dark matter injections of $e^\pm$ can also be probed through various other considerations, including observations of the 511 keV line from $e^\pm$ annihilation \cite{Wilkinson:2016gsy} as well as secondary photons \cite{Cirelli:2023tnx,Crocker:2010gy,Fermi-LAT:2015att,Abazajian:2010sq} produced by synchrotron radiation of $e^\pm$ and inverse Compton scattering of ambient photons (CMB, starlight, etc) off the high energy $e^\pm$. The null result of these searches has been used to bound DM annihilation into various channels except for the diphoton one considered here. Studies on dark matter (DM) annihilation via the diphoton channel, which then create charged particles with higher order diagrams, have been lacking because dark matter is often assumed to be electrically neutral, which leads to the standard expectation that its cross section to two photons must be radiatively induced at one loop and therefore suppressed. Obtaining these other cosmic-ray limits on $\left<\sigma v\right>_{\gamma\gamma}$ requires further dedicated work and is beyond the scope of this paper. Given the very large blob parameter space currently available for detection with VERITAS, CTA, and PANOSETI as displayed in Fig.~\ref{fig:moneyplot} we do not expect the inclusion of these other cosmic-ray limits to affect our main results significantly.

\section{Dark Blob Models}
\label{section:blobmodels}

\subsection{Burst-Producing Model: Summary}
\label{ss:modelsummary}

We consider in this subsection an example model for dark matter blobs that produce upon a collision a burst of mediator particles that subsequently decay to photons, producing a gamma-ray burst. We briefly summarize the main features of the model here. For the interested reader, a more complete description of the model is given in Appendix.~\ref{appendix:modeldetails}.

The particle content of our model includes a fermion $\chi$ as the fundamental dark matter particle and two mediators: a light scalar $\phi$ and a heavy scalar $S$. We assume that the dark matter is completely asymmetric and comes in the form of identical dark blobs: large composite bound states of fermions $\chi$ confined by attractive forces mediated by the light scalar $\phi$. We also assume that prior to a collision each blob has radiated away its internal energy to the point of degeneracy, which effectively shuts off further radiation. When two blobs collide, the relative motion of the blobs causes them to appear excited with respect to each other's Fermi sea. This enables a fraction [up to $\mathcal{O}(1)$] of their translational kinetic energy at impact to be released, in this case, via bremsstrahlung of heavy scalars $S$ which promptly decay to photons, thus producing a gamma-ray burst.

The nature of the blob collision in our model depends on the assumed coupling strength $g_S$ of the fermion $\chi$ to the heavy scalar $S$, and can be classified into two qualitatively different regimes:
\begin{itemize}
    \item \textit{Weakly dissipative collision}\\
    For a sufficiently weak $g_S$, the mean free path of $S$-mediated scatterings between $\chi$ particles during a blob collision can be larger than the blob radius. As such, the colliding blobs are for the most part transparent to each other. A small fraction of $\chi$ particles may scatter during the course of the blob collision. When they do, there is a small probability that a heavy scalar $S$ is emitted via bremsstrahlung. In this case, the energy per unit blob mass $\epsilon_*$, peak photon energy $E_*$, and duration $\Delta t_*$ of the gamma-ray burst from the decay of $S$ are set by the bremsstrahlung rate, the Fermi energy of the $\chi$ particles, and the duration of the blob collision, respectively.
    \item \textit{Strongly dissipative collision}\\
    We define this regime to be when the coupling strength $g_S$ is sufficiently strong that the mean free path of $S$-mediated $\chi$ scatterings is shorter than the blob radius. We expect the rapid $\chi$ scatterings to efficiently dissipate all of the kinetic energy of the blob's relative bulk motion into random thermal motion, causing them to stop and merge into a single object. Once the blobs' bulk kinetic energy is converted to thermal energy, it is just a matter of time for the thermal energy to be radiated away through $S$ bremsstrahlung. In this case, the resulting gamma-ray burst will have its $\epsilon_*$ set by the entire bulk kinetic energy of the blobs at impact, its $E_*$ set by the Fermi energy of the $\chi$ particles, and its $\Delta t_*$ set by the longer of the blob collision timescale and the time it takes for the merger product to radiate $\mathcal{O}(1)$ of its internal energy. 
\end{itemize}

The gamma-ray burst signal in the weakly dissipative collision regime is relatively weak but can be estimated robustly, whereas the signal in the strongly dissipative collision regime is less tractable but can be many orders of magnitude stronger. Accordingly, we find that the expected number detectable of events in VERITAS, CTA, and PANOSETI is $N_{\rm event}=\mathcal{O}\left(10^{-8}-10^{-7}\right)$ at the highest in the weakly dissipative case, while it can go as high as $N_{\rm event}=\mathcal{O}\left(10^{3}-10^{5}\right)$ and more in the strongly dissipative case. If the blobs form binaries in the early universe \cite{2017PhRvD..96l3523A,Raidal:2018bbj,Inman:2019wvr,Jedamzik:2020ypm,Raidal:2024bmm,Diamond:2021dth,Bai:2023lyf}, we find that the significantly enhanced blob collision rate can raise the expected number of detectable events in the three detectors to $N_{\rm event}=\mathcal{O}\left(0.1-10\right)$ in the weakly dissipative case.

\subsection{Formation Mechanisms}
\label{ss:formation mechanism}

The Standard Model sector has provided us with a wide variety of examples of large fermion bound states with known formation mechanisms. The dark sector should be able to form analogous bound states through similar mechanisms if it possesses certain key features of the SM. We first discuss two classes of dark blob formation scenarios that are largely based on known SM processes, namely, nucleosynthesis and star formation, albeit with simplification and tweaks. 

In \textit{dark nucleosynthesis} \cite{Hardy:2014mqa, Gresham:2017cvl, Krnjaic:2014xza} production scenarios, the formation of dark blobs proceeds in a way similar to how nuclei are synthesized during Big Bang Nucleosynthesis (BBN): heavier dark bound states are produced sequentially from lighter ones through a chain of exothermic reactions. Key ingredients in this scenario are fermions with short-range attractive interactions analogous to nucleons with nuclear forces that bind them together in a nucleus. Since, unlike in BBN, dark nucleosynthesis may be free of bottlenecks and Coulomb barriers, the synthesized fermion bound states may grow unthwarted to macroscopic sizes.

Another class of dark blob formation scenarios draws inspiration from how stars form. The formation mechanism begins with primordial dark matter density fluctuations growing gradually during matter domination until they become nonlinear, followed by the nonlinear regions decoupling from the Hubble flow, and collapsing into virialized halos. Note that the dark sector may evolve in this way up to this point without any additional interactions beyond gravity. With that said, long-range attractive self-interactions stronger than gravity can enhance the tendency of the dark sector particles to clump and have an instability similar to the gravitational collapse occur much earlier, during radiation domination instead of matter domination \cite{Savastano:2019zpr,Domenech:2023afs}. 
Finally, compact dark blobs would form if the virialized halos can efficiently cool and further contract without annihilating. The success of this scenario therefore requires the dark sector be asymmetric and equipped with dissipative interactions \cite{Chang:2018bgx,Bramante:2024pyc,Flores:2020drq}.

Large fermion bound states can also form in a class of scenarios involving a cosmological first-order phase transition from a (higher-energy) false vacuum to a (lower-energy) true vacuum \cite{Hong:2020est,Witten:1984rs,Bai:2018dxf, Gross:2021qgx, Asadi:2021pwo,Asadi:2021yml}. While first-order phase transition does not occur in the Standard Model \cite{Aoki:2006we}, it may occur in theories with a new higgs-like scalar \cite{Hong:2020est,Xie:2024mxr} as well as in those with confining gauge theories, e.g. $SU(N)$, with appropriate fermionic content \cite{Witten:1984rs, Bai:2018dxf, Gross:2021qgx, Asadi:2021pwo, Asadi:2021yml}. Independent of the first-order phase transition, the theory must include preexisting dark fermions (the constituents of the eventual dark blobs) that are energetically favored to remain in the false-vacuum phase. These dark blob formation scenarios share the following broad-brush chronology of events. Once the dark sector cools below the critical temperature of the first-order phase transition, bubbles of the true-vacuum begin to nucleate sporadically and proceed to expand. As the true-vacuum bubbles grow in both number and size, the dark fermions, unable to enter the true vacuum bubbles, are collected in the shrinking false-vacuum region between the true-vacuum bubbles. The true-vacuum region eventually percolates and occupies most of the Hubble volume. By that point, the false-vacuum region has been reduced to relatively small pockets containing trapped dark fermions. These dark fermion pockets then cool down, become increasingly compact, and gradually approach the compactness of the present epoch's blobs.

There are also other dark blob formation mechanisms that do not fall into the above categories \cite{DelGrosso:2024wmy, Balkin:2023xtr, Kusenko:1997si,Amin:2010dc}. Depending on the model and parameter space, dark blobs may leave imprints on the CMB \cite{Dvorkin:2013cea,Caloni:2021bwp}, emit gravitational waves \cite{Diamond:2021dth,Banks:2023eym,Bai:2023lyf,Barrau:2024kcb,Kuhnel:2018mlr}, produce cosmic-ray anti-nuclei \cite{Fedderke:2024zzk}, cause damage tracks in various materials \cite{Ebadi:2021cte,Bhoonah:2020dzs,Acevedo:2021tbl,Clark:2020mna}, and cause dynamical heating of stars in ultrafaint dwarfs \cite{Graham:2023unf,Graham:2024hah}, among others \cite{Jacobs:2014yca, Grabowska:2018lnd,Baum:2022duc,Du:2023dhk,Acevedo:2020avd, Bai:2019ogh,Croon:2024jhd,Mathur:2021gej,Xiao:2024qay}.

\section{Discussion}
\label{section:discussion}

The success of the wavefront technique relies on being able to discriminate gamma-ray burst signals from cosmic-ray backgrounds based on the morphology and time development of their digitized Cherenkov images. To better identify and reconstruct ultrashort gamma-ray burst events, it is useful to extend the simulations of gamma-ray-wavefront-induced showers to a wider range burst durations and individual gamma-ray energies, as well as to cases where the incoming gamma-rays have non-monochromatic spectra and large zenith angles. Besides that, to confirm our rough estimates of the burst signals in our example dark matter model, it would be interesting to perform numerical simulations of blob collisions based on the Boltzmann-Uehling-Uhlenbeck (BUU) equation. We leave these studies as future works.

Secondary charged particles produced by gamma-ray wavefronts may also be detected directly in water Cherenkov detectors, such as HAWC, LHAASO, ALPACA, ALTO, LATTES, and SGSO. Water Cherenkov detectors have the advantages of having a large FoV, a high duty cycle, and a short dead time. However, without an imaging ability it is unclear if we can distinguish gamma-ray wavefront events from single high-energy gamma-ray or cosmic ray events. In a future work, we plan to study the sensitivities of water Cherenkov detectors to ultrashort gamma-ray bursts.

Space-based gamma-ray detectors such as Fermi-LAT have the advantages of having large FoVs and duty cycles, and are essentially free of astrophysical backgrounds. If their pile-up time and dead time issues can be overcome, they would be excellent detectors for ultrashort gamma-ray bursts. One approach to effectively improve the time resolutions of space-based detectors is to use two or more of them in conjunction to look for coincident signals. Proposals for networks of space-based telescopes have been put forward (though not intended for detecting ultrashort bursts) \cite{Inceoglu:2020pos,Greiner:2022hxe}. While each detector can register at most one photon within its pile-up or dead time, simultaneous photon detection in multiple detectors within a short time interval would indicate a gamma-ray burst signal with a duration comparable or less than the time interval. A network of Fermi-LAT-like detectors operating for a decade, for example, will allow us to extend the search for rare gamma-ray bursts with occurrence rates $\gtrsim 0.1 \text{ burst/year}$ to shorter burst durations $\Delta t_*\lesssim 10\,\mu\text{s}$. Moreover, a network of space-based detectors can also be used to triangulate the distance of the burst source \cite{Ukwatta:2015mfb}.

In the dark matter model discussed in Section.~\ref{ss:modelsummary} and Appendix.~\ref{appendix:modeldetails}, the strongest burst energy per unit blob mass $\epsilon_*$ corresponds to the scenario where the entire kinetic energy of the colliding blobs at impact (which may derive from the blobs' mutual binding energy) is converted to gamma-rays. Blob collisions can, in principle, create even stronger signals, up to $\epsilon_*\approx 1$, in other models if we can tap into the mass energy of the blobs, through either decay or annihilation of the blob constituents. In these cases, one needs to explain why the decay/annihilation occurs only when two blobs are colliding and not when they are freely floating in isolation. Such a situation can be realized, for instance, if some particle-antiparticle segregation mechanism in the early universe (see e.g.~\cite{Goolsby-Cole:2015chd,Shaposhnikov:2023hrx}) leads to the formation of blobs and anti-blobs that collide and annihilate into Standard Model particles in the present epoch. Alternatively, one can consider dark matter particles that are unstable when free but stable inside a blob, similar to how neutrons behave inside and outside a neutron star. Along this line, a blob collision may eject a large number of free, unstable dark matter particles that subsequently decay to photons \cite{Bondorf:1980ahh,Morawetz:2000iu}. Yet another possibility is to have a blob collision trigger a runaway collapse of the blob merger product that leads to many orders of magnitude increase in its density \cite{Fedderke:2024zzk}. At some point, the blob's density becomes sufficiently high as to turn on some higher-dimensional operators that convert the dark matter particles of the collapsing blob merger product into Standard Model particles.

We have focused on ultrashort bursts of gamma-rays in this paper. It would be interesting to consider ultrashort bursts of photons in different energy ranges than that considered here. Sources of transient gamma-rays considered here may also produce other types of transients, e.g. non-gamma-ray photons, neutrinos, and gravitational waves. These possible counterparts to our gamma-ray signals are also interesting exploratory targets. Since the burst timescales we are interested in are too short for reorientation of detectors for multi-messenger observations, large-FoV surveys instruments would be needed to observe coincident events and identify these counterparts.

\section{Conclusion}
\label{section:conclusion}
Motivated by the prospects of probing unexplored signals that are within technological reach, we have explored methods for detecting gamma-ray transients that may occur rarely, perhaps once a year, and last only briefly, for less than ten microseconds. Because past and present gamma-ray detectors have poor sensitivities to such infrequent and short bursts, there might exist a class of astronomical, perhaps beyond the Standard Model, objects emitting such signals that have escaped detection thus far. Hence, there are opportunities for new discoveries once this observational window is opened.

We have proposed methods to search for ultrashort gamma-ray bursts based on an existing detection scheme, devised by Porter and Weekes in 1978 and developed further by the SGARFACE collaboration more than a decade ago, called the wavefront technique. While gamma-rays from conventional transient sources arrive sparsely enough to be detected individually through imaging their showers with Cherenkov telescopes, bursts that are sufficiently intense and brief can only be detected collectively since the thin wavefront of densely arriving gamma-rays initiate air showers that are highly overlapping. The wavefront technique seeks to identify ultrashort burst signals by matching the unique Cherenkov images of their superimposed air showers with the results of air shower simulations.

There are at least three ways to extend the search for ultrashort gamma-ray bursts with current and near-future experiments. First, one can search for wavefront signals in the archival data of existing IACTs, since such signals are not already searched for in the standard analysis of IACT data. Similar analyses can be done on the data that CTA will collect, with improved sensitivity and reach compared to the existing IACTs. Beside IACTs, a near-future facility called PANOSETI will survey large patches of the sky for optical transients at ten-nanosecond time-resolutions and so will be sensitive to Cherenkov lights from gamma-ray wavefronts. Further, dramatic improvements in the sensitivities of these facilities to a wide range of burst durations can be achieved by simply running a copy of the collected data through a new trigger system.  We have estimated the reaches of IACTs and PANOSETI if they were to run for a decade. Their projected sensitivities to gamma-ray bursts span many orders of magnitude in energy fluences, burst durations, and burst rates that are thus far unprobed.

While ultrashort bursts can be searched for regardless of their sources, if we assume the bursts arise from dark matter, several constraints specific to dark matter become relevant. These include those arising from the more or less known density distribution of dark matter, the Bullet Cluster, gravitational microlensing, cosmic rays, gamma-ray observations of galaxy clusters, diffuse gamma-ray background, and possibly more. We have examined the radius vs mass plane of dark matter bound states (blobs) accounting for these limits and assuming that when two blobs collide some fraction of the blobs' mass energy is released in a gamma-ray burst, but otherwise agnostic to the microphysics of the blobs. We find that more than ten orders of magnitude in the masses and radii of dark matter blobs can be probed with IACTs and PANOSETI.

We have also constructed a concrete model of large fermionic dark matter bound states that produce an ultrashort gamma-ray burst upon a collision. In this model, when two blobs collide, the fermions of the blobs scatter and emit copious amounts of mediators via bremsstrahlung. These mediators then rapidly decay to Standard Model particles, including gamma-rays, producing a gamma-ray burst. While, in general, the burst strength is determined by the coupling between the fermion and the mediator, among other considerations, up to $\mathcal{O}(1)$ of the translational kinetic energy of the colliding blobs can be radiated impulsively in this way. After accounting for model-specific constraints, our estimates suggest that gamma-ray burst signals in this model can be detected with the above mentioned techniques.

\acknowledgments
We thank Savas Dimopoulos, Peter Graham, Roni Harnik, Jamie Holder, Simon Knapen, Frank Krennrich, Alexander Kusenko, Tugdual LeBohec, Nicola Omodei, Roger Romani, and David A. Williams for useful conversations. E.H.T. thanks Michael Fedderke and Anubhav Mathur for useful discussions on a previous project. E.H.T.~acknowledges support by NSF Grant No. PHY-2310429, Simons Investigator Award No. 824870, and Gordon and Betty Moore Foundation Grant No.~GBMF7946. This work was supported by the U.S.~Department of Energy~(DOE), Office of Science, National Quantum Information Science Research Centers, Superconducting Quantum Materials and Systems Center~(SQMS) under Contract No.~DE-AC02-07CH11359. D.E.K.~and S.R.~are supported in part by the U.S.~National Science Foundation~(NSF) under Grant No.~PHY-1818899.
S.R.~is also supported by the Simons Investigator Grant No.~827042, and by the~DOE under a QuantISED grant for MAGIS. 
D.E.K.~is also supported by the Simons Investigator Grant No.~144924.

\appendix

\section{Sensitivities of IACTs and PANOSETI to Ultrashort Gamma-Ray Bursts}
\label{appendix:wavefronttechnique}

\begin{figure}
    \centering
    \includegraphics[width=\linewidth]{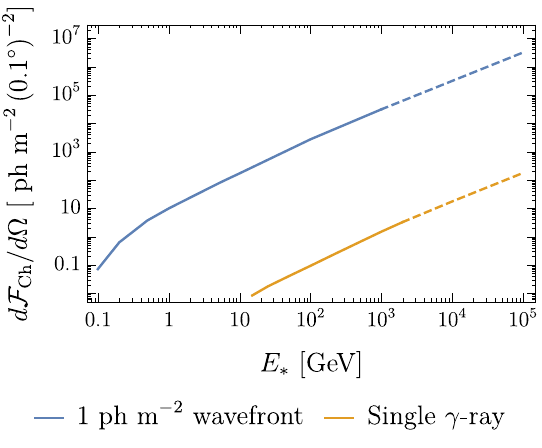}
    \caption{The number of Cherenkov photons per unit ground area per unit solid angle on the sky $d\mathcal{F}_{\rm Ch}/d\Omega$ produced in the showers initiated by an infinitely thin wavefront of gamma-rays arriving at zenith with fluence $1\text{ ph m}^{-2}$ adopted from the simulation results of \cite{LeBohec:2005wt,Schroedter:2008tp,2009arXiv0908.0182S}, plotted as a function of the energy $E_*$ of the primary gamma-rays. Also shown for comparison is the $d\mathcal{F}_{\rm Ch}/d\Omega$ due to a single gamma-ray photon at zenith, obtained from \cite{ONG199893,Arbeletche:2020rev}.  The value of $d\mathcal{F}_{\rm Ch}/d\Omega$ shown here is what is observed at $0.5^\circ$ zenith angle, which corresponds to the half width half maximum of the angular profile of $d\mathcal{F}_{\rm Ch}/d\Omega$ due to a gamma-ray wavefront. The dashed lines are the linear extrapolations of the available simulation results.}
    \label{fig:Chenrenkov}
\end{figure}

A gamma-ray burst with a delta-function pulse profile originating from a faraway source would strike the upper atmosphere in the form an infinitely thin, planar wavefront. Each of the arriving gamma-rays then produces an air shower that is spread over an area of $A_{\rm pool}\sim 10^{5}\text{ m}^2$ at ground level, known as the shower pool. For sufficiently high gamma-ray fluence, $\mathcal{F}_*\gg 1\text{ ph}/A_{\rm pool}$, a large number of showers overlap within an $A_{\rm pool}$, and the primary $\gamma$-ray photons can thus be detected \textit{collectively} through their cumulated Cherenkov yields. This method of detecting the primary gamma-rays is called the \textit{wavefront technique} \cite{1978MNRAS.183..205P, Krennrich:1999tu,LeBohec:2002hq,Schroedter:2008tp,2003ICRC....5.2971L,Schroedter:2009iw,2005APh....23..235L} and is different from the typically used technique for conventional sources that detects primary gamma-rays individually \cite{2023hxga.book..144E,Prandini:2022wcb,Bose:2021oiq}. In this appendix, we provide simple estimates for the sensitivities of IACTs and PANOSETI, specifically to gamma-ray wavefront events. The sensitivity will be expressed in terms of the minimum value of the energy fluence $E_*\mathcal{F}_*$ of the incoming wavefront, i.e. the minimum number of gamma-rays per unit ground area $\mathcal{F}_*$ multiplied by the energy of individual gamma-rays $E_*$, for a detection.

The Cherenkov light from multi-photon-initiated showers was simulated and shown to be detectable with IACTs \cite{Krennrich:1999tu,LeBohec:2002hq,Schroedter:2008tp}. According to the simulations, a simultaneous wavefront of $E_*\gtrsim 10\GeV$ gamma-ray primaries with fluence $\mathcal{F}_{*}$ produces Cherenkov photons at ground level with a typical fluence\footnote{We choose to show the Cherenkov fluence per solid angle of $(0.1^\circ)^2$, which is significantly smaller than a square degree, because the Cherenkov fluence varies considerably within and does not extend beyond a square degree. Apart from that this choice is arbitrary and is meant to reflect the typical FoV of a PMT used in IACTs.} 
\begin{align}
    \frac{d\mathcal{F}_{\rm Ch}}{d\Omega}\approx 
    300
    \text{ ph}/\text{m}^2/\left(0.1^\circ\right)^2\left(\frac{E_*\mathcal{F}_{*}}{10\GeV/\text{m}^2}\right)
\end{align}
over a time spread of $\mathcal{O}(10\text{ ns})$ and an angular spread $\theta_{\rm Ch}\sim 1^\circ$. Note that the above value of $d\mathcal{F_{\rm Ch}}/d\Omega$ is half of the maximum value, which occurs at $\approx 0.5^\circ$ zenith angle. The approximate linear dependence of  $d\mathcal{F}_{\rm Ch}/d\Omega$ on $E_*$ breaks at $E_*\lesssim 10\GeV$. See Fig.~\ref{fig:Chenrenkov} for the complete $E_*$ dependence.

Cherenkov photons from a gamma-ray burst would form a compact image of triggered pixels on IACT or PANOSETI cameras. These images are then analyzed in order to reconstruct the properties of the primary particles producing the shower. The reconstruction is based on Monte Carlo simulations of multi-photon-initiated air showers together with the responses of the detector. The results of the simulations are compiled into a multi-dimensional table, called a look-up table, that maps the properties of the primary gamma-ray burst (photon energy, fluence, incoming direction, and more) to the parameters of the digitized shower images (centroid position, orientation, size, shape, and more). Given the parameters of a shower image, one can invert the table to determine the properties of the primary particles.

\subsection{VERITAS}
Detector properties vary considerably between telescopes. We take VERITAS as a reference \cite{2011ICRC....9...14K,Schroedter:2009iw}. VERITAS is an array of four optical telescopes, where each telescope consists of a mirror with large collecting area $A_{\rm mirror}=95\text{ m}^2$ to focus Cherenkov lights from secondary shower particles toward $N_{\rm PMT}=499$ photomultiplier tubes (PMTs). Each PMT employed by VERITAS records photons from a patch of the sky with a solid angle of radius $\theta_{\rm PMT}=0.15^\circ$, corresponding to a solid angle $\Omega_{\rm PMT}=2\pi\left(1-\cos(\theta_{\rm PMT}/2)\right)=(0.13^\circ)^2$, and converts them to electrons, with an efficiency $\eta_{\gamma\rightarrow e}=0.19$ \cite{2011ICRC....9...14K,2011arXiv1110.4702N} (this is the new number after the PMTs were upgraded; the old value before the upgrade was $\eta_{\gamma\rightarrow e}=0.14$ \cite{Schroedter:2009iw}). In our definition, $\eta_{\gamma\rightarrow e}$ includes both the mirror reflectivity and the PMT quantum efficiency. 
Assuming an integration time window just enough to catch the entire shower duration and neglecting the angular dependence of the collecting area of the mirror, the number of photoelectrons registered by a PMT whose FoV is within the $\theta_{\rm shower}\sim 1^\circ$ shower spread is given by
\begin{align}
    N_{\text{ph}e}\sim\frac{d\mathcal{F}_{\rm Ch}}{d\Omega}\Omega_{\rm PMT} A_{\rm mirror}\eta_{\gamma\rightarrow e}\nonumber\
\end{align}
Requiring this be greater than the trigger threshold of $N_{\text{ph}e}^{\rm thr}=5$ \cite{Schroedter:2009iw} implies a minimum gamma-ray energy fluence for triggering an event
\begin{align}
    E_*\mathcal{F}_{\rm det}\approx &\,5\times 10^{-3}\GeV/\text{m}^2 \left(\frac{N_{\text{ph}e}^{\rm thr}}{5}\right)^{-1}\nonumber\\
    &\times \left(\frac{\Omega_{\rm PMT}}{\left(0.13^\circ\right)^2}\right)^{-1}\left(\frac{A_{\rm mirror}}{95\text{ m}^2}\right)^{-1}\left(\frac{\eta_{\gamma\rightarrow e}}{0.19}\right)^{-1}\label{eq:sensitivity}
\end{align}
for $E_*\gtrsim 10 \GeV$. The result for the full energy range is shown in Fig.~\ref{fig:fluencesensitivity}. When we set the photon to photoelectron conversion efficiency to the old value $\eta_{\gamma\rightarrow e}=0.14$ used in \cite{Schroedter:2009iw}, we find that $\mathcal{F}_{\rm det}\approx 3.6\times 10^{-2}\text{ ph}/\text{m}^2$ at $E_*=1 \GeV$, which roughly agrees with the fluence sensitivity of VERITAS to a 10 ns burst reported in \cite{Schroedter:2009iw}. We collect these assumed detector properties in Table~\ref{tab:my_label}.

The range of primary gamma-ray energy for which IACTs are sensitive to is often quoted to be $100\GeV\lesssim E_\gamma\lesssim 10\TeV$ \cite{2023hxga.book..144E,Prandini:2022wcb,Bose:2021oiq}. The lower bound is set by the number of Cherenkov photons produced in a gamma-ray induced shower that can be detected with a $\sim 100\text{ m}^2$ IACT mirror, which drops rapidly below $\sim 100\GeV$. The upper bound is set by the uncertainty in the energy reconstruction from Cherenkov image analysis which becomes $\mathcal{O}(1)$ or worse at $E_\gamma\gtrsim 10\TeV$. We note that these apply for single gamma-ray events but not necessarily for gamma-ray wavefront events in which we are interested. The wavefront technique has been shown to be effective for $100\MeV\lesssim E_*\lesssim 100\GeV$ \cite{Krennrich:1999tu,LeBohec:2002hq,Schroedter:2008tp,2003ICRC....5.2971L,Schroedter:2009iw,2005APh....23..235L}. Further simulations will be needed to determine the applicability of the wavefront technique at higher $E_*$.

In the main text, we show results assuming energy fluence sensitivities that could be achieved if all backgrounds can be rejected. For completeness, we briefly discuss the most important backgrounds:
\begin{itemize}
\item \textit{Night sky background (NSB).}\\
The flux of the NSB is approximately $d\Phi_{\rm NSB}/d\Omega\sim 10^{12}\text{ ph}/\text{m}^2/\text{s}/\text{sr}$ which, in terms of flux, can be a significant background for the Cherenkov photon signals, even when considering a 10 ns window. However, the NSB occurs at random and so tends to trigger PMTs in a sporadic way. False triggers from the NSB can be eliminated by requiring several neighboring pixels to be triggered simultaneously. 

\item \textit{Cosmic-ray-induced showers.}\\
The superimposed Cherenkov yields from a wavefront of gamma-rays have distinctive angular and temporal distributions from that of cosmic-ray showers. Compared to most cosmic-ray events, multi-photon-initiated showers have images that are larger, more circular, and also smoother (superposing a large number of gamma-ray initiated showers averages out shower fluctuations). These differences can be exploited to reject cosmic-ray backgrounds. Additionally, the images from a wavefront of gamma-rays appear essentially the same in different telescopes, whereas cosmic-ray-induced showers have different shapes and orientations in different detectors. By cross-correlating events detected in different telescopes in search for coincident and similar events, one can further reject background events. With enough telescopes, the telescope array collectively can perhaps be made essentially background-free.
\item \textit{Isotropic diffuse gamma-ray background (IGRB).}\\
Fermi has measured the IGRB flux to be
$\sim 0.3\text{ ph/m}^2/\text{s}/\text{sr}$ at $100\MeV$ and lower at higher photon energies \cite{Fermi-LAT:2014ryh}. The background showers initiated by these individual gamma-rays will again yield Cherenkov images different from that due to multi-photon-induced shower signals, allowing us to reject them similar to how cosmic-ray events are rejected.
\end{itemize}
To reject these backgrounds, VERITAS adopts the following three-level trigger algorithm. The Level 1 trigger operates at the single-pixel level, requiring a single PMT trigger threshold of 5 photo-electrons. The Level 2 trigger checks for coincident triggers of multiple pixels: at least three adjacent PMTs need to exceed the threshold within a $5\text{ ns}$ window. Finally, at Level 3 at least two telescopes must be triggered within a 50 ns coincidence window. In projecting the sensitivity of other telescopes, we ignore these complications and simply assume that these backgrounds can be rejected completely using similar trigger algorithms.

\begin{table}
    \centering
    \caption{Detector properties. Area stands for either the mirror area (VERITAS and CTA) or the Fresnel lens area (PANOSETI).}
    \begin{tabular}{|c|c|c|c|c|}
    \hline
     &VERITAS&CTA&PANOSETI\\
     &(1 telescope)& (1 LST)& (1 module)\\\hline
     $\theta_\text{FoV} [^\circ]$ &$3.5$&$4.3$&$9.92$\\\hline
     $\theta_{\rm PMT} [^\circ]$&$0.15$&$0.1$&$0.31$\\\hline
     No. PMTs&499&1855&1024\\\hline
     Area [m$^2$]&95&370&0.166\\\hline
     Peak $\eta$&0.19\footnote{upgraded PMT from \cite{2011ICRC....9...14K,2011arXiv1110.4702N}}&0.42&0.25\\\hline
\end{tabular}
    \label{tab:my_label}
\end{table}

\subsection{CTA}

CTA is the next-generation IACTs that will succeed the current-generation ones: VERITAS, MAGIC, and HESS \cite{Hofmann:2023fsn}. CTA is planned to have 3 sets of telescopes arrays: Large-Sized Telescopes (LSTs), Medium-Sized Telescopes (MSTs), and Small-Sized Telescopes (SSTs). In projecting the sensitivity of CTA, we adopt the properties of LSTs which have the largest photon collection area among the three sets. LSTs consist of four telescopes, each with a mirror of area $A_{\rm mirror}=370\text{ m}^2$ and $N_{\rm PMT}=1855$ photomultiplier tubes (PMTs) at its focus. The PMTs used in the LSTs have an FoV angular diameter of $\theta_{\rm PMT}=0.1^\circ$, corresponding to a solid angle $\Omega_{\rm PMT}=2\pi\left(1-\cos(\theta_{\rm PMT}/2)\right)=(0.089^\circ)^2$, and a photon to photoelectron conversion efficiency of $\eta_{\gamma\rightarrow e}=0.42$. These relevant properties of CTA can be found in Table.~\ref{tab:my_label}

\subsection{PANOSETI}

Since its first proposal, the potential configurations of PANOSETI have been updated frequently. For the purpose of deriving its projected sensitivity, we adopt one of it latest proposed configurations as outlined in \cite{Maire:2021etj}, specifically the one called Configuration A. In this configuration, PANOSETI is a geodesic dome with 80 modules. These modules function as independent telescopes pointing in different directions, which collectively cover a large part of the sky. Unlike IACTs, which use mirrors,
PANOSETI telescopes use Fresnel lenses to collect and focus light. Each PANOSETI module consists of a $0.166\text{ m}^2$ area lens that focuses light onto a $32\times32$ pixel photon counting detector. Here, each pixel is a silicon photomultiplier (SiPM) instead of a conventional PMT used in IACTs.\footnote{While SiPM suffers from dark current and noise from cross-communication between cells, and requires more complex algorithm \cite{Ruch2014}, the trigger criterion of an SiPM is similar to that of a conventional PMT, but with a trigger threshold of $N_{\rm phe}^{\rm thr}=11.5$ \cite{2022SPIE12184E..8BM,wright2018panoramic}.} Each SiPM employed by PANOSETI records photons from a square-shaped patch of the sky with a side length $\theta_{\rm SiPM}=0.31^\circ$, corresponding to a solid angle $\Omega_{\rm SiPM}=(0.31^\circ)^2$, and converts them to photoelectrons, with an efficiency $\eta_{\gamma\rightarrow e}=0.25$ \cite{Maire:2021etj,Korzoun:2023jgb}. Adding up all $32\times 32=1024$ pixels in a telescope results in a field of view of $\Omega_{\rm FoV}=(9.92^\circ)^2$. Since there are 80 identical modules within a dome, the total field of view of our chosen PANOSETI configuration is $\Omega_{\rm SiPM}^{\rm tot}=2.27$ sr, excluding the redundant coverage \cite{Maire:2021etj}.

We note that some of the modules in the PANOSETI dome are pointing at large zenith angles. Simulations of the showers produced by gamma-ray wavefronts from large zenith angles have not been reported. The shower profiles in those cases may be qualitatively different from that at small zenith angles since the amount of air traversed by shower particles is larger at higher zenith angles. Dedicated simulation studies are thus needed to obtain an accurate estimate for the sensitivity of large-FoV instruments like PANOSETI. In this paper, we simply assume that Eq.~\eqref{eq:sensitivity} applies and substitute into it the properties of PANOSETI as described above and outlined in Table~\ref{tab:my_label}.

\section{Fireball}
\label{appendix:fireball}

An injection of a large amount of Standard Model particles in a small volume of space over a short duration of time may lead to the formation of a \textit{fireball} \cite{1986ApJ...308L..47G, 1986ApJ...308L..43P,Piran:1999kx,Meszaros:2006rc}, an optically thick plasma that self-accelerates to ultrarelativistic bulk radial velocities under its own radiation pressure. In this appendix, we discuss the condition for fireball formation and how fireball formation affects the properties of the perceived gamma-ray bursts.

\subsection{Thermalization}
Consider a sudden and spatially localized injection of Standard Model particles with a total energy $\epsilon_{\rm SM} M$ and no net quantum number of any kind. The condition for these particles to thermalize depends on the details of the injection. Here, we discuss as an example the case where the injection is entirely in the form of photons, such that the total energy injected in the form of photons is $\epsilon_{\gamma,\rm inj.}M=\epsilon_{\rm SM}M$. We further assume that the photon injection is concentrated in a spherical volume of radius $R_{\rm inj.}$, lasts for a duration of $t_{\rm inj.}$, and has a monochromatic spectrum with energy $E_{\rm inj.}$. These input parameters $\epsilon_{\gamma,\rm inj.}$, $R_{\rm inj.}$, $t_{\rm inj.}$, and $E_{\rm inj.}$ of the injection are to be distinguished from the output parameters $\epsilon_*$, $\Delta t_*$, and $E_*$ of the burst used in the main text. 

In order for the injected particles to reach a chemical equilibrium, number-changing reactions are required \cite{Mukaida:2015ria,Harigaya:2013vwa,Davidson:2000er, Diamond:2023scc,Chang:2022gcs,Garny:2018grs}. As a rough estimate of the thermalization condition, we consider only the number-changing process $ee\rightarrow ee\gamma$, where $e$ without a superscript stands for either an electron $e^-$ or a positron $e^+$. We assume that thermalization can be achieved if the  $ee\rightarrow ee\gamma$ process is efficient inside the injection radius $R_{\rm inj.}$.\footnote{Outside the injection radius, $r\gtrsim R_{\rm inj}$, the efficiency of a process scales approximately as $n\sigma v_{\rm rel}r\propto r^{-1}$ or faster, since the number density of the relevant particles scales as $n\propto r^{-2}$ \cite{Fedderke:2024zzk,MacGibbon:2007yq}. The cross section $\sigma$ can be assumed independent of $r$ for particles that have not interacted since their production. The relative velocity of the particles participating in the process $v_{\rm rel}$ may decrease with $r$, which is why $n\sigma v_{\rm rel}r$ may decrease with increasing $r$ faster than $r^{-1}$.} Since this process involves electrons and/or positrons, we first need to estimate the amount of $e^\pm$ produced by the injected photons. The efficiency of $\gamma\gamma\rightarrow e^+ e^-$ within the injection radius $r\lesssim R_{\rm inj.}$ is
\begin{align}
    \Gamma_{\gamma\gamma\rightarrow e^+e^-}R_{\rm inj.}\sim &~n_{\gamma,\rm inj.}\sigma_{\gamma\gamma\rightarrow e^+e^-}R_{\rm inj.}\nonumber\\
    \sim &~4\times 10^3\times \text{ min}\left[1,\frac{R_{\rm inj.}}{\Delta t_{\rm inj.}}\right]\left(\frac{\epsilon_{\rm SM} M}{10^{23}\text{ g}}\right)\nonumber\\
    &~\times \left(\frac{E_{\rm inj.}}{1\text{ TeV}}\right)^{-3}\left(\frac{R_{\rm inj.}}{1\text{ ns}}\right)^{-2}
\end{align}
In writing the second line, we assumed $\sigma_{\gamma\gamma\rightarrow e^+e^-}\sim \pi\alpha^2/E_{\rm inj.}^2$ which applies for $E_{\rm inj.}\gg m_e$ and
\begin{align}
    n_{\gamma,\rm inj.}\sim \text{ min}\left[1,\frac{R_{\rm inj.}}{\Delta t_{\rm inj.}}\right]\frac{\epsilon_{\rm SM} M}{R_{\rm inj.}^3E_{\rm inj.}}
\end{align}
where we included a factor $\text{min}\left[1,R_{\rm inj.}/\Delta t_{\rm inj.}\right]$ to account for burst ($\Delta t_{\rm inj.}\lesssim R_{\rm inj.}$) and wind ($\Delta t_{\rm inj.}\gtrsim R_{\rm inj.}$) branching of cases \cite{Fedderke:2024zzk}. In the wind case, $n_{\gamma,\rm inj.}$ is suppressed by a factor of $R_{\rm inj.}/\Delta t_{\rm inj.}\lesssim 1$ because only a fraction $R_{\rm inj.}/\Delta t_{\rm inj.}$ of the total number of injected photons is inside the injection region at a given time. If $\Gamma_{\gamma\gamma\rightarrow e^+e^-}R_{\rm inj.}$ turns out to be greater than unity then it means $e^+e^-\leftrightarrow \gamma\gamma$ will reach a detailed balance, in which case $n_e\sim n_{\gamma,\rm inj.}$. Accounting for both possibilities, namely $\Gamma_{\gamma\gamma\rightarrow e^+e^-}R_{\rm inj.}\lesssim 1$ and $\Gamma_{\gamma\gamma\rightarrow e^+e^-}R_{\rm inj.}\gtrsim 1$, the initial number density of electrons produced from $\gamma\gamma\rightarrow e^+e^-$ can be written as 
\begin{align}
    n_{e}\sim n_{\gamma,\rm inj.}\text{min}\left[1,\Gamma_{\gamma\gamma\rightarrow e^+e^-}R_{\rm inj.}\right]
\end{align}
These electrons and positrons can subsequently undergo $ee\rightarrow ee\gamma$ with the rate \cite{Chang:2022gcs,Garny:2018grs}
\begin{align}
    \Gamma_{ee\rightarrow ee\gamma}\sim f_{\rm dead cone}\times \alpha\times \text{min}\left[\Gamma_{ee\rightarrow ee}^{\rm soft},t_{\rm form}^{-1}\right]
\end{align}
where $\Gamma_{ee\rightarrow ee}^{\rm soft}\sim \alpha^2n_e/\omega_p^2$ is the infrared-divergent electron collision rate regulated by the Debye scale, $t_{\rm form}^{-1}\sim \text{max}\left[m_e^2/E_{\rm inj.}, \alpha^{1/2}\omega_p\right]$ is the inverse timescale for the produced photons to separate from the electron that produces it (which accounts for the Landau–Pomeranchuk–Migdal effect) \cite{Arnold:2001ba,Kurkela:2011ti}, $f_{\rm dead cone}\sim\text{min}\left[1,(\omega_p/m_e)^2\right]$ is a factor included to account for the dead cone effect \cite{Dokshitzer:2001zm}, and $\omega_p\sim \sqrt{\alpha n_e/E_{\rm inj.}}$ is the order of magnitude of the plasma frequency of the initial, non-thermal $e^\pm$ plasma \cite{Carrington:1997sq,Arnold:2002zm}. 

For example, for the set of parameters $M=10^{23}\text{ g}$, $R_{\rm inj.}\sim t_{\rm inj.}\sim 1\text{ ns}$, and $E_{\rm inj.}=1\TeV$,  
we find $\Gamma_{ee\rightarrow ee\gamma}R_{\rm inj.}\sim 2\times 10^8$ if $\epsilon_{\rm SM}=1$ and $\Gamma_{ee\rightarrow ee\gamma}R_{\rm inj.}\sim 5\times 10^{-5}$ if $\epsilon_{\rm SM}=10^{-6}$. These indicate that the rough condition for fireball formation, $\Gamma_{ee\rightarrow ee\gamma}R_{\rm inj.}\gtrsim 1$, can be satisfied in at least some part of the injection parameter space ($\epsilon_{\gamma,\rm inj.}$, $R_{\rm inj.}$, $t_{\rm inj.}$, $E_{\rm inj.}$). Nevertheless, as we will discuss below, the burst parametrization and the results we obtained in the main text remain valid whether or not a fireball forms following an injection from a blob collision.

\subsection{Output}
Once the fireball achieves thermalization and attains a semirelativistic bulk radial velocity $v\sim 1$, its subsequent evolution is dictated by hydrodynamics \cite{Chang:2022gcs,Fiorillo:2023ytr,Fiorillo:2023cas} and can be parametrized with its initial temperature $T_0$ and initial radius $R_0$. The initial fireball temperature $T_0$ can be obtained from energy conservation $\epsilon_{\rm SM} M\times \text{min}\left[1,R_{\rm inj.}/\Delta t_{\rm inj.}\right]=g_{*,0}(\pi^2/15)T_0^4(4\pi R_{\rm inj.}^3)$ while the initial fireball radius $R_0$ is set by the injection radius $R_{\rm inj.}$
\begin{align}
    T_0\approx &~0.8\GeV\times \text{min}\left[1,\frac{R_{\rm inj.}}{\Delta t_{\rm inj.}}\right]^{1/4}\left(\frac{g_*}{12}\right)^{-1/4}\nonumber\\
    &~\times \left(\frac{\epsilon_{\rm SM}M}{10^{23}\text{ g}}\right)^{1/4}\left(\frac{R_{\rm inj.}}{1\text{ ns}}\right)^{-3/4}\label{eq:T0}\\
    R_0\sim & ~R_{\rm inj.}
\end{align}
where $g_*=12$ corresponds to $\gamma,e^\pm, \mu^\pm, \pi^{0,\pm}$. As long as the energy density of the fireball is dominated by radiation, the comoving temperature $T'$ and bulk Lorentz factor $\gamma$ of the fireball evolves as
\begin{align}
    T'&\approx T_0\frac{R_0}{R}\\
    \gamma&\approx \frac{R}{R_0}
\end{align}
In what follows, primed quantities are defined in the comoving frame of the fireball. Note that these scalings preserve the average energy per particle in the center-of-mass frame of the fireball, $\gamma T'\approx T_0=\text{constant}$. Hence, when the photons are released from the fireball, their energy spectrum will peak at $E_*\sim T_0$. The duration of the burst will be set the thickness of the plasma shell when it becomes optically thin. In the absence of baryon loading \cite{1986ApJ...308L..47G, 1986ApJ...308L..43P,1990ApJ...365L..55S,Grimsrud:1998me,2008ApJ...677..425L}, the fireball expansion preserves the plasma shell thickness $\Delta R\approx R_0\sim R_{\rm inj.}$ in the burst case $\Delta t_{\rm inj.}\lesssim R_{\rm inj.}$ or the wind length $\Delta R\approx \Delta t_{\rm inj.}$ in the wind case $\Delta t_{\rm inj.}\gtrsim R_{\rm inj.}$. Thus, the burst duration of the released gamma-rays will be given by $\Delta t_*\sim \text{max}\left[\Delta t_{\rm inj.}, R_{\rm inj.}\right]$.

We now estimate the comoving temperature at which the fireball plasma becomes optically-thin. Consider the process $e^+e^-\leftrightarrow \gamma\gamma$ which initially keeps the $e^\pm$ number density at its chemical equilibrium value
\begin{align}
    \left(n_e^\prime\right)_{\rm ch}\approx \left(\frac{m_e T'}{2\pi}\right)^{3/2}e^{-m_e/T'}
\end{align}
where we have for simplicity assumed that the $e^\pm$ are non-relativistic because the decoupling most likely occurs after Boltzmann suppression has kicked in. The $e^+e^-$ annihilation ceases to be efficient when the annihilation rate $\Gamma_{e^+e^-\rightarrow\gamma\gamma}^\prime \sim \left(n_e^\prime\right)_{\rm ch}(8\pi/3)(\alpha^2/m_e^2)$ becomes slower than the comoving expansion rate of the fireball $\gamma R^{-1}\sim R_0^{-1}$, i.e. when
\begin{align}
    \left(\frac{m_e T'}{2\pi}\right)^{3/2}e^{-m_e/T'}\frac{8\pi \alpha}{3m_e^2}\sim R_0^{-1}
\end{align}
Solving this for $T'$, we find that electron-positron annihilation stops being efficient when the comoving temperature is
\begin{align}
    T'_{e^+e^-\nrightarrow \gamma\gamma}=\mathcal{O}(10 \keV)
\end{align}
with some logarithmic dependence on $R_0$ \cite{1990ApJ...365L..55S,Grimsrud:1998me,2008ApJ...677..425L}. 

Since $T'_{e^+e^-\nrightarrow \gamma\gamma}\ll m_e$, by the time the $n_e^\prime$ freezes out the electron-positron number density $n_e^\prime$ is severely Boltzmann suppressed compared to $n_\gamma^\prime\sim T_{e^+e^-\nrightarrow \gamma\gamma}^{\prime 3}$ by many orders of magnitude. It follows that the vast majority of the particles released from the fireball are photons, which means $\epsilon_*\approx \epsilon_{\rm SM}$. After the $e^\pm$ density has frozen out, though most $e^\pm$ can scatter with the more abundant photons, most of the photons do not scatter with electrons or positrons anymore since the Thomson scattering rate from the point of view of the photons is of the same order as $\Gamma_{e^+e^-\rightarrow\gamma\gamma}^\prime$. In other words, the properties of gamma-ray photons that escape the bulk of the fireball are essentially the same as those at $e^+e^-\leftrightarrow \gamma\gamma$ decoupling.

We therefore find that in the case where the injected Standard Model particles subsequently thermalize into a fireball, the properties of the resulting gamma-ray burst are
\begin{align}
    \epsilon_*&\approx \epsilon_{\rm SM}\\
    E_*&\sim  T_0\\
    \Delta t_*&\sim \text{max}\left[\Delta t_{\rm inj.},R_{\rm inj.}\right]
\end{align}
with $T_0$ given by Eq.~\eqref{eq:T0}. The energy spectrum of the burst is that of a boosted thermal distribution such that the peak energy is $E_*\sim T_0$. Note that $E_*$ and $\Delta t_*$ can be arbitrarily large or small depending on the details of the initial injections: $\epsilon_{\rm SM}M$, $R_{\rm inj.}$ and $\Delta t_{\rm inj.}$.

To summarize, if a fireball forms following an injection of Standard Model particles then the $\epsilon_*$ of the resulting gamma-ray burst will be driven close to $\epsilon_{\rm SM}$, while the $E_*$ and $\Delta t_*$ can still take up essentially any values depending on the input parameters $\epsilon_{\rm SM}M$, $R_{\rm inj.}$, and $\Delta t_{\rm inj.}$. In other words, fireball formation tends to increase $\epsilon_*$ and does not significantly limit the possible values of
$E_*$ and $\Delta t_*$. These imply not only that fireball formation does not affect our main analysis (which parametrizes bursts in terms of the output parameters $\epsilon_*$, $E_*$, and $\Delta t_*$), but also that it provides further motivation to assume most of the injected Standard Model energy ends up in gamma-rays.

\section{Burst-Producing Model: Details}
\label{appendix:modeldetails}
The particle content of our model includes a fermion $\chi$ with mass $m_\chi$ which couples to a light scalar $\phi$ with mass $m_\phi$ and a heavy scalar $S$ with mass $m_S\gg m_\phi$. The Lagrangian of the model reads
\begin{align}
    \mathcal{L}\supset &\bar{\chi}(i\slashed{\partial}+m_\chi)\chi-g_\phi\phi\bar{\chi}\chi+\frac{1}{2}\left(\partial \phi\right)^2-\frac{1}{2}m_\phi^2\phi^2\nonumber\\
    &-g_S S\bar{\chi}\chi+\frac{1}{2}\left(\partial S\right)^2-\frac{1}{2}m_S^2S^2-\frac{g_{S\gamma\gamma}}{4}SF^2
\end{align}
where $F$ is the electromagnetic field strength tensor (with Lorentz indices omitted), the fermions $\chi$ are the fundamental dark matter particles, the light scalar field $\phi$ mediates long-range attractive forces that bind $\chi$ particles together into blobs, and the heavy scalars $S$ are to be emitted during a blob collision before subsequently decaying into photons (and potentially other light Standard Model particles), producing a gamma-ray burst.

\subsection{Fermi Degenerate Blobs}
We assume that the dark matter particle $\chi$ is completely asymmetric \cite{Kaplan:2009ag} and exists in its entirety in the form of identical blobs of mass $M$ and radius $R$. Each blob consists of $N_\chi=M/m_\chi$ fermions $\chi$ bound together by attractive forces mediated by the light scalar $\phi$ whose range is longer than the size of the blob, $m_\phi^{-1}\gg R$,\footnote{Blobs bound together by short-range Yukawa forces are also viable. The motivation behind the choice $m_\phi^{-1}\gtrsim R$ is that it increases both the blob collision rate and the blob velocities prior to an impact. As we will see, the latter enhances the amount of energy available for radiation and ameliorates Pauli blocking as blob collision is taking place.} but short enough that constraints on dark matter self-interactions at Galactic scales and beyond (e.g.~\cite{Bogorad:2023wzn}) do not apply. The heavy scalar $S$ also mediates attractive forces between $\chi$ particles. In the parameter space we consider, where $m_S^{-1}\ll R$ and $g_S\gg g_\phi$, the $S$-mediated attraction between $\chi$ particles affect the structure of the blob negligibly but is the dominant force that facilitates hard collisions between $\chi$ particles during a blob collision. Specifically, the condition that the pressure contribution from $S$-mediated attraction is negligible compared to that from $\phi$-mediated attraction boils down to $g_S^2\ll m_S^2/(m_\chi^2 v_F)$, which turns out to be a subdominant constraint on $g_S$ in our analysis below.

We also assume that the fermions $\chi$ in each blob are degenerate, i.e. their temperature is small compared to their Fermi energy. We expect the internal evolution of a blob to naturally bring itself toward degeneracy. Although the details of this process will depend on the formation mechanism of the blob and so are beyond the scope of this work, we can make several qualitative remarks. Depending on the temperature of the blob, the blob may lose energy predominantly via either on-shell $S$ emissions or $S$-mediated photon emissions. If the blob begins in a thermal-pressure supported hydrostatic equilibrium, which has a negative heat capacity, the energy loss would initially heat the blob and at the same time cause it to shrink in volume. In that case, the blob only begins to cool when it has shrunk sufficiently that it switches from being thermal-pressure supported to degeneracy-pressure supported, whereupon its heat capacity becomes positive. The cooling then shuts off asymptotically as the momentum states available for scattering are gradually restricted by Pauli exclusion to the progressively thinner vicinity of the Fermi sphere.

Furthermore, we assume for simplicity that the fermions $\chi$ are uniformly distributed inside each blob with a mass density $\rho_\chi=M/(4\pi R^3/3)$, a number density $n_\chi=\rho_\chi/m_\chi$, and a Fermi momentum $p_F=(3\pi^2 n_\chi)^{1/3}$, which we assume to be non-relativistic, $p_F\lesssim m_\chi$. Hydrostatic equilibrium between $\phi$-mediated attraction and Fermi repulsion implies the following radius-mass relation for the blobs \cite{Wise:2014jva}
\begin{align}
    R\approx \frac{60}{g_\phi^2M^{1/3}m_\chi^{2/3}}\label{eq:RMrelation}
\end{align}
It follows that the Fermi velocity of the fermions $\chi$ in a blob is 
\begin{align}
    v_F=\frac{p_F}{m_\chi}\approx 5\times 10^{-2} g_\phi^2\left(\frac{M}{m_\chi}\right)^{2/3}
\end{align}
which is a key quantity for our analysis below. Note that the escape velocity of a $\chi$ particle out of the blob is comparable to its Fermi velocity, $v_{\rm esc}\sim v_F$.

\subsection{Blob Formation}
\label{ss:formation}

While the blobs described in the previous subsection can form in any number of ways, a simple paradigm is that these blobs result from the collapse of large $\chi$ density perturbations in the early universe, reminiscent of a widely studied scenario for PBH formation \cite{Zeldovich:1967lct,Hawking:1971ei,Carr:1974nx}. When the density contrasts of $\chi$ hit $\mathcal{O}(1)$, their dynamics become nonlinear, and they start to collapse. It is often the case that the collapsing nonlinear structures virialize before reaching their Schwarzschild radii and halt the collapse \cite{Musco:2018rwt,Musco:2020jjb,Savastano:2019zpr,DeLuca:2021pls,Padilla:2021zgm,Harada:2022xjp}. The virialized halos can contract further as they radiate away their internal kinetic energies, until their Fermi pressure becomes significant and stops the contraction. In our model, the latter occurs at a radius far greater than the Schwarzschild radius, ensuring that the halos relax into stable blobs instead of turning into PBHs. Therefore, generally speaking, mechanisms proposed to form PBHs can be repurposed as blob formation mechanisms under suitable choice of parameters \cite{Hong:2020est,Flores:2020drq,Flores:2022oef}.

In this paradigm, the blobs' masses are set by the comoving sizes of the density fluctuations from which they form. Their radii are, in turn, set by the radius-mass relation, as given in Eq.~\eqref{eq:RMrelation}. Many early universe scenarios predict enhanced primordial curvature perturbations on small scales \cite{Ozsoy:2023ryl,Green:2024bam,Villanueva-Domingo:2021spv,Inomata:2024dbr,Ji:2021mvg}. When a sufficiently large perturbation reenters the horizon, it collapses immediately, resulting in a blob whose mass is comparable to the horizon mass of $\chi$ at that time. The blobs' abundance and mass function in this case thus depend on the details of inflationary physics responsible for the enhanced primordial curvature perturbations. In particular, we expect the blob's mass function to be peaked where the corresponding curvature perturbations is the largest. Furthermore, since the blobs' energy density scales like matter, its energy density fraction can grow substantially during radiation domination. The present-epoch DM abundance can be achieved even if large overdensities are rare and/or the blobs are highly subdominant initially.

Alternatively, perturbations in the density of $\chi$ can start small, and grow to large values via an instability, before they collapse upon reaching a threshold. Such an instability occurs naturally in the presence of long-range attractive forces stronger than gravity \cite{Savastano:2019zpr,Domenech:2023afs,Flores:2020drq, Flores:2022oef}, whose role can be played by the light scalar field $\phi$ in our model. Shortly after the onset of the instability, the overdensities of $\chi$ become nonlinear and proceed to collapse. The masses of the resulting blobs in this case are approximately monochromatic, given roughly by the $\chi$ mass within the fifth force range $m_\phi^{-1}$ at that time or when $m_\phi^{-1}$ first becomes subhorizon \cite{Flores:2020drq, Flores:2022oef}. Although we require $m_\phi^{-1}$ to be significantly greater than the radius of the present-epoch blobs $R$ in our main analysis, this is a very weak requirement. Thus, $m_\phi^{-1}$ is essentially a free parameter as far as blob formation is concerned, and can be tuned to yield blobs of virtually any mass. Additionally, introducing a quartic coupling of $\phi$ provides another knob with which we can tune the blob's mass \cite{Domenech:2023afs}. This scenario predicts that $\mathcal{O}(1)$ of $\chi$ particles will reside in blobs. The overall abundance of the $\chi$ particles itself can be regarded as an initial condition provided by, e.g., inflationary physics \cite{Flores:2020drq, Flores:2022oef}.

We have seen that the blobs' mass and abundance depend on not only unconstrained parameters of our model, but also on additional parameters tangential to the model. Hence, blob formation considerations do not necessarily impose additional constraints on the model parameters that are relevant for observational signals in the present epoch, such as mass and abundance of the blobs. Accordingly, we leave them as free parameters for the rest of our analysis. The scenarios discussed above by no means exhaust all possible ways to form blobs; see Section.~\ref{ss:formation mechanism} for alternatives.

\subsection{Sommerfeld Enhancement}
Consider two blobs approaching each other with the typical virial velocity in the Galaxy, $v_{\rm DM}\sim 10^{-3}$. Once their relative distance falls within the range $m_\phi^{-1}$ of the $\phi$ mediator, $\phi$-mediated attraction between the blobs turns on and accelerates the blobs toward each other. At the moment of impact, the kinetic energy of the blobs, $Mv_*^2$, is their kinetic energy at infinity, $M v_{\rm DM}^2$, plus the binding energy between the blobs
\begin{align}
    Mv_*^2=Mv_{\rm DM}^2+\frac{g_\phi^2N_\chi^2}{4\pi (2R)}
\end{align}
The radius-mass relation \eqref{eq:RMrelation} of the blob implies that the second term is $\sim Mv_F^2$. Hence, the velocity of the blobs right before the impact is
\begin{align}
    v_*\sim \text{max}\left[v_{\rm DM},v_F\right]    
\end{align}
In the interest of producing a strong burst signal, the case where $v_F\gg v_{\rm DM}$ (but $v_F$ not ultrarelativistic) is more interesting,\footnote{In the opposite case, $v_F\lesssim v_{\rm DM}$, the scalar $S$ needs to be sufficiently light ($m_S\lesssim m_\chi v_{\rm DM}^2$) in order to be emitted, to the point that strong stellar cooling and BBN limits on $g_S$ become relevant.} and this is what we will assume in our subsequent analysis. Furthermore, the attractive force between the blobs leads to a Sommerfeld enhancement of the velocity-averaged blob collision cross section
\begin{align}
    \left<\sigma v\right>_{\rm col}\sim \left(\frac{v_F}{v_{\rm DM}}\right)^2\times v_{\rm DM}R^2
\end{align}
as long as $m_\phi^{-1}\gtrsim (v_F/v_{\rm DM})R$.

\subsection{Weakly Dissipative Collision}

We first consider in this subsection the simplest case where, when two blobs collide, the majority of the fermions $\chi$ do not participate in scatterings and the blobs simply pass through each other to first approximation. We refer to this case as \textit{weakly dissipative collision}. The $S$ emission in this case is easier to model, though the resulting gamma-ray burst signal is relatively weak. In the next subsection~\ref{ss:stronglydissipative}, we consider the opposite, \textit{strongly dissipative collision}, case where most $\chi$ fermions in the colliding blobs scatter multiple times during the course of a blob collision. Although the blob collision process is less tractable in the strongly dissipative case, we expect the resulting gamma-ray burst signal to be much stronger.

Before going into the details, we sketch here the basic picture we have in mind for the weakly dissipative collision case. In a blob collision, which we assume to be always head-on, the two blobs simply pass through each other to zeroth order, i.e., $\propto g_S^0$ in $g_S$ expansion. The spatial and momentum distributions of the fermions may change during the collision due to $\mathcal{O}(1)$ changes in the mean field $\left<\phi\right>$, but not drastically. A small $\propto g_S^4$ fraction of $\chi$ particles participate in hard $\chi\chi\rightarrow\chi\chi$ scatterings mediated by the heavy scalar $S$, and when they do there is a small $\propto g_S^2$ probability that an $S$ particle is emitted via bremsstrahlung $\chi\chi\rightarrow \chi\chi S$ with $S$ exchange in the $t$ channel, which implies a $\propto g_S^6$ emission rate of $S$. The produced $S$ particles then free stream away from the collision region and promptly decay to gamma-rays. Thus, in this regime, the typical energy of the emitted gamma-rays is set by the Fermi energy of the $\chi$ particles $E_*\sim E_F$, the minimum burst duration is set by the blob overlap time $\Delta t_*\sim R/v_*\sim R/v_F$, and in the limiting case where $\mathcal{O}(1)$ of $\chi$ particles scatter in a collision the burst energy per unit blob mass is given by $\epsilon_*\sim 2g_S^2v_F^6/(15\pi^2)$. We elaborate on these in what follows.

\subsubsection{$\propto g_S^0$: Pauli exclusion}

We start by neglecting collisions between $\chi$ particles altogether. This corresponds to taking the limit of vanishing $g_S$. The fermion dynamics in this limit is driven mainly by the effects of the mean field $\left<\phi\right>$ and Pauli exclusion. As mentioned previously, we are mainly interested in the regime where the blobs' relative velocity at impact $v_*$ is comparable to the Fermi velocity of each blob. An important question to address is how Pauli exclusion manifests as two degenerate blobs with overlapping Fermi spheres begin overlapping spatially.

Free fermions move in the $(\mathbf{x},\mathbf{p})$ phase space along straight lines that do not cross one another. Two free fermions that are initially detached will evolve in such a way that they will never overlap. If at an instance the momentum of fermion A is such that it would occupy an adjacent infinitesimal phase-space patch occupied by fermion B, the time evolution will automatically be such that fermion B (which must have the same momentum as fermion A in order to be momentum degenerate with it) leaves the patch as fermion A enters it.  

When constraints are involved, however, the way Pauli exclusion manifests is less trivial. A blob of degenerate fermions is best thought of as a gas of essentially free fermions, with long mean free paths, contained in the blob by the mean field $\left<\phi\right>$ \cite{feldmeier1986dissipative,feldmeier1987transport}. The long mean free path of the fermions can be attributed to Pauli blocking of most collisions due to the would-be final states being occupied. The gradient of the mean field $\left<\phi\right>$ acts as a confining vessel for the $\chi$ particles. Right before an impact, the Fermi spheres characterizing the occupation numbers $f(\mathbf{p})$ of the $\chi$ particles in the blobs are displaced from one another by $m_\chi v_*$. Since, $v_*\sim v_F$, there is generically considerable overlap between the two Fermi spheres. When the blobs first touch, the fermion distributions of the two blobs overlap in the momentum space but not yet in the configuration space. At this point, it is clear that the two blobs do not overlap in phase space.

It is interesting to ask what happens to the fermions as two blobs with overlapping Fermi spheres begin to also overlap in the configuration space. While quantum Boltzmann (BUU) simulations may be required to determine the subsequent phase-space evolution of the fermions \cite{Kruse:1985pg, Buck:1983lyf,Biro:1987kob,Bauer:1986zz,Aichelin:1985zz}, we expect that Pauli exclusion will be obeyed through rearrangements in the momentum space: the fermion distribution in the momentum space will swell in order to avoid double occupancy as blob interpenetration is taking place. As the interpenetration of the blobs progresses, the volume of the compound system formed by the two blobs decreases. Since Pauli exclusion keeps the fermion's phase-space density constant\footnote{According to Liouville's theorem, fermions behave collectively as an incompressible fluid in the $(\vec{x},\vec{p})$ phase space, i.e., it may change in shape but always keeps its phase-space volume (or phase-space density). This property ensures the Pauli exclusion principle is respected at all times. } a spatial squeezing must be accompanied by a momentum-space (i.e. Fermi sphere) swelling. Analogous effects are found in the context of nuclear collisions \cite{Trefz:1985ag,Beck:1978gg,PhysRevC.64.014601}. The swelling of the momentum-space distribution can also be thought of as being the result of the moving boundaries of the colliding blob system, reminiscent of Fermi acceleration.\footnote{Classically one expects an elastic collision of a particle with a wall moving toward the particle with velocity $v_w$ to increase the velocity of the particle by $2v_w$. Similar effect occurs in the quantum case, as can be seen by studying a toy problem of fermions in a potential well with one of the walls moving inward \cite{cooney2017infinite,cassing1987quantal,Dittrich:2024obz}. In that case, one would find that the expectation value of the kinetic energy of the particle $\left<E\right>$ increases with time.}

There are several $\mathcal{O}(1)$ effects that can speed up or slow down the blobs during the course of a blob collision. Yukawa attraction between blobs tends to cause the relative velocity of the blobs to speed up during the first half of the collision and slow down during the second half. An opposite tendency arises from the increase in the internal Fermi kinetic energy of $\chi$ particles as a result of their momentum-space rearrangement in order to satisfy Pauli exclusion. Additionally, particle transfer between blobs may contribute to slowing down the relative velocity of the blobs \cite{feldmeier1986dissipative, feldmeier1987transport,Gross:1984xoo, randrup1979theory,Luke:1993zz,1978AnPhy.113..330B,Sierk:1980zza}. Further, it is possible that some $\chi$ particles are ejected from the blobs during the collision. These effects are difficult to quantify and we do not expect their inclusion to change our conclusions significantly. To keep the analysis tractable, we will assume in what follows that the bulk of the blobs simply pass through each other with constant bulk velocities.

\subsubsection{$\propto g_S^4$: elastic $\chi$ scattering}

Next, we turn on $\propto g_S^4$ $S$-mediated collisions between $\chi$ particles without considering  $\propto g_S^6$ bremsstrahlung processes. While kinematics dictates that all scatterings between $\chi$ particles belonging to the same degenerate blob are Pauli blocked, two $\chi$ particles can collide if they originate from different blobs. The relative bulk motion of the blobs opens up accessible phase space for the final states of $\chi$: from the perspective of a blob (the reference blob) $\chi$ particles in the other blob appear as excited states with kinetic energies $\sim m_\chi v_Fv_{\rm *}$ in excess above the Fermi energy of the reference blob.

Consider a single $\chi$ particle originating from blob 1 moving through the Fermi sea of blob 2. Such an inter-blob particle transfer is allowed for particles in the non-overlapping parts of the Fermi spheres; these transfers also create holes in blob 1, which allow collisions within blob 1 to repopulate the holes. The $\chi$ particle might eventually find a $\chi$ particle from blob 2 to scatter with. Even in this case, the majority of $\chi$ scatterings are Pauli blocked if the magnitude of the momentum transfer $\Delta p_\chi$ is soft $\Delta p_\chi\ll p_F$. However, given that the 
Fermi spheres of the two blobs are displaced by $m_\chi v_*\sim p_F$, a significant fraction of $\chi$ scatterings could lead to unoccupied final states if the momentum transfer is hard, $\Delta p_\chi\sim p_F$. In Appendix.~\ref{appendix:elasticscaterring}, we compute the $S$-mediated $\chi\chi\rightarrow\chi\chi$ scattering rate $\Gamma_{\rm el}$ while the blob collision is taking place in the limit of non-relativistic Fermi momentum, $p_F\ll m_\chi$, accounting for $\mathcal{O}(1)$ factors as accurately as practically possible. The result is
\begin{align}
    \Gamma_{\rm el}\sim \frac{g_S^4m_\chi}{48\pi^3} \label{eq:Gammaelmain}
\end{align}
Assuming that most of the $\chi$ particles do not participate in hard collisions amounts to the condition
\begin{align}
    \Gamma_{\rm el}\frac{R}{v_F}\lesssim 1 \label{eq:rarechicol}
\end{align}
This assumption allows us to neglect the effect of $\chi\chi\rightarrow\chi\chi$ scatterings on the relative bulk motion of the colliding blobs.

\subsubsection{$\propto g_S^6$: bremsstrahlung of heavy scalar $S$}

We now consider the $S$ production during a blob collision via $\propto g_S^6$ bremsstrahlung processes $\chi\chi\rightarrow\chi\chi S$ with $S$ exchange in the $t$ channel. The $S$ luminosity per unit of logarithmic energy $\omega$ interval of the heavy scalar $dL_S/d\ln\omega$ emitted during a blob collision can be computed to be
\begin{align}
    \frac{dL_S}{d\ln \omega}\sim \frac{4g_S^2 v_F^4}{15\pi^2}\times N_\chi\omega\Gamma_{\rm el} \label{eq:dLSdlnw}
\end{align}
The details of our computation of $dL_S/d\ln\omega$ can be found in Appendix.~\ref{appendix:bremsstrahlung}. Here, we summarize the key assumptions that go into our calculation of $L_S$. We work in the so-called soft bremsstrahlung limit, where the recoil of the emitting fermion pairs due to $S$ emission is vanishingly small compared to the momentum transfer of the $\chi$ collision $\Delta p_\chi$. We utilize the fact that the amplitude for the bremsstrahlung process $\chi\chi\rightarrow\chi\chi S$ in this limit factorizes into an elastic scattering part and an $S$ emission part, as per Low's theorem \cite{Low:1958sn,Rrapaj:2015wgs}. This assumption is justified because, for non-relativistic $\chi$ particles, $E_F\ll p_F$, and so the energy of the emitted heavy scalar $\omega\lesssim E_F$ is small compared to the typical $\chi$ momentum transfer in a hard scattering, $\Delta p_\chi\sim p_F$ in the bulk of the allowed phase space for the $\chi\chi\rightarrow\chi\chi S$ process. Moreover, we note that the leading bremsstrahlung emission from a pair of identical fundamental particles is quadrupolar instead of dipolar. Finally, we found that finite density effects on the dispersion relation of $S$ is negligible.

If the $dL_S/d\ln\omega$ found in Eq.~\eqref{eq:dLSdlnw} were to be preserved, the produced $S$ particles must free stream out of the colliding blob system unscathed. In other words, the optical depth of the colliding blobs to $S$ particles must be much smaller than unity. Effects that contribute to the optical depth of $S$ include Compton scattering $\chi S\rightarrow \chi S$ and inverse bremsstrahlung $\chi\chi S\rightarrow \chi\chi$. These effects lead to optical depth contributions of order $\tau_{\chi S\rightarrow\chi S}\sim n_\chi(g_S^4/m_\chi^2)R(E_F/p_F)\sim g_S^4v_F^4m_\chi R$ and $\tau_{\chi\chi S\rightarrow\chi\chi}\sim g_S^6v_Fm_\chi R$, respectively. We found that Compton scattering dominates the optical depth of $S$. Nevertheless, the condition $\tau_{\chi S\rightarrow\chi S}\ll 1$, after accounting for tractable $\mathcal{O}(1)$ factors, amounts to $g_S^4v_F^4m_\chi R\ll 72 \pi^3$ and is less restrictive compared to Eq.~\eqref{eq:rarechicol}.

The $dL_S/d\ln\omega$ found in Eq.~\eqref{eq:dLSdlnw} grows linearly with $\omega$, suggesting that it is the strongest at the highest possible $\omega$ allowed by kinematics, namely $\omega \sim E_F$. 
However, at around that point, the soft bremsstrahlung approximation and the resulting $dL_S/d\ln\omega\propto \omega$ behavior begin to break down. Kinematics implies that the $dL_S/d\ln\omega$ should instead turn over and decrease with $\omega$. Nonetheless, our soft bremsstrahlung approximation should still be $\mathcal{O}(1)$ correct there, and we expect the emission of $S$ to be peaked at $\omega\sim E_F$, with a magnitude not far from the prediction of Eq.~\eqref{eq:dLSdlnw}. The detailed time evolution of the $S$ emission is not important for our analysis. Integrating $dL_S/d\ln\omega$ over $\omega$ and the $\sim R/v_F$ duration of a blob collision, we find the total energy released in $S$ particles to be
\begin{align}
    L_S\frac{R}{v_F}\sim \frac{4g_S^2 v_F^4}{15\pi^2}\times \frac{1}{2}Mv_F^2\times \Gamma_{\rm el}\frac{R}{v_F} \label{eq:timeintegratedluminosity}
\end{align}
The strongest signal in the weakly dissipative regime occurs when $\Gamma_{\rm el}$ saturates Eq.~\eqref{eq:rarechicol}, $\Gamma_{\rm el}R/v_F\sim 1$. In this limiting case, the decay of $S$ particles will result in a gamma-ray burst with
\begin{align}
    \epsilon_*&\sim \frac{2g_S^2v_F^6}{15\pi^2}\label{eq:epsilonstarweak}\\
    E_*&\sim \frac{1}{2}m_\chi v_F^2\\
    \Delta t_*&\sim \text{max}\left[\frac{R}{v_F},\left(\frac{E_*}{m_S}\right)\tau_S\right]
\end{align}
where $\epsilon_*$ is obtained by dividing Eq.~\ref{eq:timeintegratedluminosity} with the blob mass $M$ and $\tau_S=(g_{S\gamma\gamma}^2m_S^3/64\pi)^{-1}$ is the decay lifetime of $S$ in its rest frame.

\subsection{Strongly Dissipative Collision}
\label{ss:stronglydissipative}
We have previously considered the weakly dissipative regime where most $\chi$ particles do not participate in scatterings during a blob collision, $\Gamma_{\rm el}(R/v_F)\lesssim 1$, in which case the colliding blobs simply overlap temporarily before they separate. We expect that stronger signals can be achieved in the same model in the strongly dissipative regime, $\Gamma_{\rm el}(R/v_F)\gtrsim 1$, where the previous treatment breaks down. While dedicated numerical simulations might be required to accurately determine the non-trivial dynamics of the blob collision in this regime, we provide a rough estimate of the resulting gamma-ray burst signal here.

When two blobs collide in this case, we expect the frequent $\chi\chi\rightarrow\chi\chi$ scatterings to efficiently dissipate the bulk kinetic energy of the blobs' relative motion and turn it into internal kinetic energy corresponding to random motions of $\chi$. At the same time, this causes the blob to stop and merge into one \cite{Samaddar:1985zz,PhysRevC.37.1959,PhysRevC.64.014601,Nakayama:1986zz}.\footnote{Some of the $\chi$ particles may be ejected from the colliding blob system. Given that the blobs' kinetic energy at impact is comparable to its potential energy associated with $\phi$-mediated attraction, we expect this to be an $\mathcal{O}(1)$ effect at most.} Since eventually all the internal thermal energy of $\chi$ will be radiated away through $\chi\chi\rightarrow\chi\chi S$ bremsstrahlung, the burst energy per blob mass in this case is given by the incoming bulk kinetic energy of the blobs
\begin{align}
    \epsilon_*\sim\frac{M v_*^2}{M}\sim v_F^2 \label{eq:epsilonstarstrong}
\end{align}
The remaining question is how long it will take to radiate this worth of energy away. This will depend on a number of factors which we will discuss below.

While thermalization is taking place, Pauli blocking amounts to an $\mathcal{O}(1)$ suppression in the bremsstrahlung rate of $S$. As thermalization is near its completion, the Fermi sphere turns approximately spherical and its sharp edges get smeared out by thermal fluctuations. By energy conservation, the temperature of the blob at this point should be of order $m_\chi v_F^2$ which is comparable to the Fermi energy of the blob. So, Pauli blocking remains to be an $\mathcal{O}(1)$ effect right after thermalization and the $S$ bremsstrahlung luminosity $L_S$ is still given approximately by Eq.~\eqref{eq:dLSdlnw}. Given this luminosity, the time it takes to radiate away the thermal energy is $t_{\rm cool}\sim M v_F^2/L_S$. As the blob radiates $S$, it initially shrinks and heats up. Once the blob has shrunk by $\mathcal{O}(1)$, the blob's Fermi pressure, which is initially comparable to its thermal pressure, becomes the dominant pressure, since the Fermi pressure grows faster than the thermal pressure with decreasing radius of the blob. Following that, subsequent $S$ emission causes blob cooling instead of heating. At the same time, the $S$ bremsstrahlung luminosity $L_S$ becomes increasingly suppressed due to Pauli blocking.

The duration of the resulting gamma-ray burst $\Delta t_*$ will depend on not only $t_{\rm cool}\sim M v_F^2/L_S$ but also the hydrodynamical timescale of the blob merger product, which in this case is set by the sound wave crossing time of the blob $t_{\rm dyn}=R/v_F$. This timescale sets both the duration of the blob collision and the time it takes for the merger product to establish a hydrodynamical equilibrium. In the limit $t_{\rm cool}\gg t_{\rm dyn}$, the blob first establishes a hydrodynamical equilibrium within $t_{\rm dyn}$, followed by a longer cooling period, where the evolution of the blob is mainly dictated by $S$ radiation. In this $t_{\rm cool}\gg t_{\rm dyn}$ case, the timescale to radiate $\mathcal{O}(1)$ of the blob's thermal energy is $t_{\rm cool}$, i.e. we have $\Delta t_*\sim t_{\rm cool}$. In the limit of highly efficient $S$ bremsstrahlung, $t_{\rm cool}\ll t_{\rm dyn}$, accessible lower-energy $\chi$ states are rapidly filled in until most attempted $\chi$ collisions are Pauli blocked and the resulting bremsstrahlung luminosity $L_S$ becomes sufficiently suppressed that $t_{\rm cool}$ exceeds $t_{\rm dyn}$. Any extra thermal energy converted from the binding energy of the blobs as they merge into a single object gets radiated away rapidly in a similar way. With this understanding, we expect the burst duration in this $t_{\rm cool}\ll t_{\rm dyn}$ case to be set by the collision timescale $t_{\rm dyn}$ itself. Therefore, accounting for both $t_{\rm cool}\gg t_{\rm dyn}$ and $t_{\rm cool}\ll t_{\rm dyn}$ cases, the $S$ burst duration can be written as
\begin{align}
    \Delta t_*\sim \text{max}\left[t_{\rm cool},t_{\rm dyn}\right]\sim \text{max}\left[\frac{M v_F^2}{L_S},\frac{R}{v_F}\right] \label{eq:Deltatstrong}
\end{align}

\subsection{Ultrashort Gamma-Ray Burst Signal}
To obtain a gamma-ray burst signal in this model, we assume that the produced mediators $S$ subsequently decay to Standard Model particles. Though there are other possibilities, it is simplest to assume that the $S$ decays directly into two photons, $S\rightarrow\gamma\gamma$.\footnote{We have assumed in our analysis so far that $S$ is a real, $CP$-even scalar. There are other possibilities. For example, the role of $S$ could be played by a pseudoscalar $a$ with an axial coupling to electromagnetism \cite{Alexander:2016aln,Battaglieri:2017aum,Chang:2023quo, Nguyen:2023czp,Luo:2023cxo,Allen:2024ndv} or by a dark photon kinetically mixed with photons \cite{DeRocco:2019njg, Diamond:2021ekg, Gan:2023wnp, Maselli:2019ubs}. In the latter case, the $\chi$ particles would need to be charged under the dark photon, thus making them millicharged particles \cite{Berlin:2022hmt, Gan:2023jbs, Mathur:2020aqv}.} We focus on the heavy scalar $S$ mass range of $100\MeV \lesssim m_S\lesssim \TeV$, for which the existing constraints on $g_{S\gamma\gamma}$ are weak, $g_{S\gamma\gamma}\lesssim 10^{-4}\GeV^{-1}$, set by searches at OPAL and ATLAS experiments \cite{Knapen:2016moh,ATLAS:2020hii}. At smaller $S$ masses, $m_S\lesssim 100\MeV$, beam dump experiment, supernova 1987A, and big bang nucleosynthesis considerations set stringent limits of $g_{S\gamma\gamma}\lesssim 10^{-11}\GeV^{-1}$ \cite{Depta:2020wmr}. Numerically, the time-dilated decay lifetime of $S$ is 
\begin{align}
    \tau_S\frac{E_*}{m_S}\sim & ~10^{-4}\text{ ns}\left(\frac{g_{S\gamma\gamma}}{10^{-4}\GeV^{-1}}\right)^{-2}\left(\frac{m_S}{\GeV}\right)^{-4}\nonumber\\
    &~\times \left(\frac{E_*}{10\GeV}\right)
\end{align}
which can easily be made to be $\lesssim R/v_{\rm F}$ such that the burst duration $\Delta t_*$ is set by the blob collision timescale $R/v_{\rm F}$ or other longer timescales.

\subsubsection{Weakly dissipative collision: free-blob}

Using Eqs.~\eqref{eq:dmax},~\eqref{eq:Nevent},~\eqref{eq:Jcol}, and~\eqref{eq:epsilonstarweak} we can compute the expected number of detectable bursts from blob collisions $N_{\rm event}$. Assuming the signal falls in the weak burst regime and accounting for the Sommerfeld enhancement of the blob collision cross section $R^2\rightarrow (v_F/v_{\rm DM})^2R^2$, we find that the number of detectable collision events scales as $N_{\rm event}\propto  M^{-1/12}E_*^{-29/12}v_F^{46/3}$. Note that this scaling ignores the $E_*$ dependence of the energy fluence sensitivity $E_*\mathcal{F}_{\rm def}(E_*)$. For $M=10^{10}\text{ g}$, $E_*=0.1\GeV$, and $v_F=0.7$, the expected number of events in each detector is
\begin{align}
    N_{\rm event}\approx \begin{cases}
        1\times 10^{-8}, &\text{VERITAS}\\
        2\times 10^{-7},&\text{CTA}\\
        2\times 10^{-8}, &\text{PANOSETI}
    \end{cases}
\end{align}

where we have assumed detector properties (FoVs $\Omega_{\rm FoV}$ and effective observation times $T_{\rm obs}$) as outlined in Section~\ref{ss:GalacticTransients}. Since we have chosen the model parameters so as to maximize $N_{\rm event}$ within the regime of applicability of our approximations, it appears to be highly unlikely that gamma-ray bursts in this case will be detected in these detectors. Note that the $M$, $E_*$, and $v_F$ assumed here correspond to $g_S=6.3\times 10^{-3}$ (which saturates Eq.~\eqref{eq:rarechicol}), $m_\chi\sim 2E_*/v_F^2=0.4\GeV$, $R=3.2\times 10^{-2}\text{ cm}$, $\Delta t_*\sim R/v_F=2\times 10^{-3}\text{ ns}$, and $\epsilon_*=6.3\times 10^{-8}$.

\subsubsection{Strongly dissipative collision: free-blob}

We have assumed in the above description that the heavy mediator luminosity $L_S$ translates directly to the energy loss of the blob which, in turn, is equal to the resulting gamma-ray burst luminosity. This is the case if the radiated particles can freely escape the blob without significant scattering or being absorbed. The condition that the blob be optically thin to $S$ particles is parametrically the same as $\Gamma_{\rm el}(R/v_F)\lesssim 1$ up to some powers of $v_F$ and other $\mathcal{O}(1)$ factors. Thus, for $v_F\sim 1$ (for which the burst signal is maximized), it is difficult to keep the blob optically thin to $S$ in the highly collisional $\Gamma_{\rm el}(R/v_F)\gtrsim 1$ regime.

However, this difficulty can be overcome by making the $S\rightarrow\gamma\gamma$ decay occur rapidly in a timescale less than the mean free time of $S\chi\rightarrow S\chi$ scattering in the blob. Given an $E_*$ in the range of our interest, the latter is easily satisfied for a wide range of $g_{S\gamma\gamma}$ well below the current experimental limit $g_{S\gamma\gamma}\lesssim 10^{-4}\GeV^{-1}$, for $m_S$ in the range $100\MeV\lesssim m_S\lesssim \TeV$, by making $m_S$ sufficiently heavy. The gamma-ray photons produced in $S$ decays may still undergo $S$-mediated $\gamma\chi\rightarrow\gamma\chi$ scattering in the blob. Requiring the blob be optically thin to these photons amounts to an upper limit on $g_{S\gamma\gamma}$ for a given $g_S$. This implies a lower limit on the decay lifetime of $S$ which, in turn, puts an upper limit on $g_S$ such that the $S\chi\rightarrow S\chi$ mean free time is longer than the decay lifetime of $S$. We find that the aforementioned limit on $g_S$ still leaves plenty of room for the burst duration $\Delta t_*$ (as given by Eq.~\eqref{eq:Deltatstrong}) to be sub-$10\,\mu\text{s}$ or even sub-$10\text{ ns}$. We have also checked that other constraints on the coupling strength $g_S$, such as those from requiring the blobs' profiles be unaffected by $S$-mediated attraction and inverse bremsstrahlung $\chi\chi S \rightarrow \chi\chi$ be negligible are less stringent.

Ignoring the $E_*$ dependence of $E_*\mathcal{F}_{\rm det}(E_*)$, we find using Eqs.~\eqref{eq:dmax},~\eqref{eq:Nevent},~\eqref{eq:Jcol}, and~\eqref{eq:epsilonstarstrong} that the expected number of observable events scale as $N_{\rm event}\propto M^{1/6}E_*^{-8/3}v_F^{25/3}$ in the weak burst regime ($d_{\rm max}\lesssim 10 \text{ kpc}$) and as $N_{\rm event}\propto M^{-4/3}E_*^{-8/3}v_F^{16/3}$ in the strong burst regime ($d_{\rm max}\gtrsim 10 \text{ kpc}$). For detector properties (FoVs $\Omega_{\rm FoV}$ and effective observation times $T_{\rm obs}$) as outlined in Section~\ref{ss:GalacticTransients} and
the set of parameters $M=10^{17}\text{ g}$, $E_*=0.1\GeV$, $v_F=0.7$, the value of $N_{\rm event}$ turns out to be
\begin{align}
    N_{\rm event}\sim \begin{cases}
        3\times 10^3, &\text{VERITAS}\\
        6\times 10^4, &\text{CTA}\\
        8\times 10^3, &\text{PANOSETI}
    \end{cases}
\end{align}
The above chosen parameters correspond to $m_\chi\sim 2E_*/v_F^2= 0.4\GeV$, $R \approx 7 \text{ cm}$, $g_S \approx 5\times 10^{-3}$, $\Delta t_*\sim R/v_F= 0.3\text{ ns}$, and $\epsilon_*\sim v_F^2=0.5$. Setting $g_{S\gamma\gamma}$ to $7 \times  10^{-5} \text{ GeV}^{-1}$ or values around it, for example, would ensure both the produced $S$ particles decay within the $S\chi\rightarrow S\chi$ scattering mean free time and the blob is optically thin to the gamma rays produced in $S$ decays. 

While our estimate for $N_{\rm event}$ is very crude, the extremely large values of $N_{\rm event}$ given by this estimate suggest that a more realistic calculation will most likely predict $N_{\rm event}\gtrsim 1$ at and around the above chosen point in the blob parameter space.

\subsubsection{Weakly dissipative collision: binary-blob}
Pairs of blobs can form binary systems in the early universe, which results in significant enhancements in their collision rate today. Consider the simplest case where the mediator $\phi$ is sufficiently short range (but still longer than the size of the blob) that the binary evolution is driven purely by gravity. In that case, the local merger rate at the present epoch was found in \cite{2017PhRvD..96l3523A} to be 
\begin{align}
    \frac{d\dot{N}_{\rm merger}}{dV}\approx 6\times 10^{-7}\text{ pc}^{-3}\text{yr}^{-1}\left(\frac{M}{10^{23}\text{ g}}\right)^{-32/37} \label{eq: mergerrate}
\end{align}
The number of detectable merger events can be estimated as $N_{\rm event}\approx d\dot{N}_{\rm merger}/dV\times \Omega_{\rm FoV}T_{\rm obs}\times d_{\rm max}^3/3$. Neglecting the $E_*$ dependence of the energy fluence sensitivity $E_*\mathcal{F}_{\rm def}(E_*)$, and using Eqs.~\eqref{eq:dmax} and \eqref{eq:epsilonstarweak}, we find that it scales as $N_{\rm event}\propto M^{57/148}E_*^{1/4}v_F^{10}$. For the set of parameters $M=10^{23}\text{ g}$, $E_*=10\GeV$, and $v_F=0.7$, the expected number of detectable events is 
\begin{align}
    N_{\rm event}\approx \begin{cases}
        0.2, &\text{VERITAS}\\
        6,&\text{CTA}\\
        1, &\text{PANOSETI}
    \end{cases}
\end{align}
assuming detector properties (FoVs $\Omega_{\rm FoV}$ and effective observation times $T_{\rm obs}$) as outlined in Section~\ref{ss:GalacticTransients}. Here $g_S=7.6\times 10^{-4}$, $m_\chi\sim 2E_*/v_F^2=40\GeV$, $R=1.5\text{ cm}$, $\Delta t_*\sim R/v_F=7\times 10^{-2}\text{ ns}$, and $\epsilon_*=9.3\times 10^{-10}$. The results indicate that if the blobs form binaries, the bursts are potentially detectable in all the detectors we consider even in the weakly dissipative case. We expect orders of magnitude higher $N_{\rm event}$ in the strongly dissipative, binary-blob case.

There are, however, several caveats to this result that may worsen and improve the detection prospect. First, while the signals from individual bursts are now stronger because of the better likelihood of blob collisions happening nearby, the IGRB limit discussed in Section~\ref{ss:IGRB} also gets stronger and so needs to be re-evaluated. There may also be additional limits from considerations of energy injections in the early universe because, unlike in the case of free-blob collisions, binary-blob mergers may occur at considerable rates in the early universe. Further, the merger rate we assumed in Eq.~\eqref{eq: mergerrate} does not account for the clustering and three-body dynamics of blobs, which may either increase or decrease the merger rate \cite{Raidal:2018bbj,Inman:2019wvr,Jedamzik:2020ypm,Raidal:2024bmm}. Moreover, strictly speaking Eq.~\eqref{eq: mergerrate} applies only for primordial black hole binaries. It may not be accurate for blobs that are sufficiently large in size compared to their Schwarzschild radii \cite{Diamond:2021dth}. Finally, we have not considered the case where the $\phi$-mediated forces are sufficiently long range so as to dominate the binary formation and dynamics \cite{Bai:2023lyf}.

\section{Rate Calculations for Burst-Producing Model}
\label{appendix:ratecalculations}
In this appendix, we derive the elastic scattering rate $\Gamma_{\rm el}$ and bremsstrahlung luminosity spectrum $dL_S/d\ln k$ used in Appendix.~\ref{appendix:modeldetails}. We denote 4-vectors with a capital letter (e.g., $P$), 3-vector with a bold lower-case letter (e.g., $\mathbf{p}$), 3-vector magnitudes with a normal lowercase letter (e.g., $p=|\mathbf{p}|$), unit 3-vector direction with a lowercase letter with a hat on it (e.g., $\hat{p}=\mathbf{p}/p$); we denote lab-frame and center-of-mass frame quantities without (e.g. $\theta$) and with a prime (e.g., $\theta^\prime$), respectively.

\subsection{Elastic Scattering $\chi\chi\rightarrow\chi\chi$}
\label{appendix:elasticscaterring}
During a blob collision, some of the $\chi$ particles undergo elastic scatterings
$\chi(P_1)\chi(P_2)\rightarrow\chi(P_3)\chi(P_4)$, with $P_i=(E_i,\mathbf{p}_i)$ being the 4-momenta of the participating $\chi$ particles. The rate at which elastic scattering occurs for a particle with a 3-momentum $\mathbf{p}_1$ is given by \cite{peskin2018introduction}
\begin{widetext}
    \begin{align}
    \Gamma_{\rm el}(\mathbf{p}_1)=\frac{1}{2}\frac{1}{2m_\chi}\int_{p_2<\tilde{p}_F(\hat{p}_2)} d\Pi_2\int  d\Pi_3 d\Pi_4 F  \sum_{\rm spins}\left|\mathcal{M}_{\rm el}\right|^2 (2\pi)^4\delta^{(4)}\left(P_1+P_2-P_3-P_4\right)
\end{align}    
\end{widetext}
where 
\begin{align}
    d\Pi_i&=\frac{1}{2E_i}\frac{d^3\mathbf{p}_i}{(2\pi)^3}\\
    F=&\left[1-f(\mathbf{p}_3)\right]\left[1-f(\mathbf{p}_4)\right]\nonumber\\
    =&\Theta\left[p_3-\tilde{p}_F(\hat{p}_3)\right]\Theta\left[p_4-\tilde{p}_F(\hat{p}_4)\right]
\end{align}
To account for possible distortions of the Fermi sphere while a blob collision is taking place, we treat the Fermi momentum as a quantity $\tilde{p}_F(\hat{p})$ that depends on the momentum direction $\hat{p}$ and also on time (though in our notation we do not show the time dependence explicitly). The value of $F$ is 1 for events with Pauli-allowed final states and 0 otherwise. Using the 3-momentum- and energy-conserving delta functions to evaluate $\int d\Pi_4$ and $\int dp_3$ integrals, respectively, we find 
    \begin{align}
        \Gamma_{\rm el}(\mathbf{p}_1)=&\frac{1}{8\pi (2m_\chi)}\int_{p_2<\tilde{p}_F(\hat{p}_2)}d\Pi_2\frac{p_3}{E_1+E_2}\nonumber\\
        &\times \int \frac{d\Omega_3}{4\pi}  F  \frac{ \sum_{\rm spins}|\mathcal{M}_{\rm el}|^2}{\left|1-\frac{|\mathbf{p}_1+\mathbf{p}_2|E_3\cos\theta_3}{p_3(E_1+E_2)}\right|}\nonumber\\
        =&\frac{1}{8\pi (2m_\chi)^3}\int_{p_2<\tilde{p}_F(\hat{p}_2)}\frac{d^3p_2}{(2\pi)^3}p_3\nonumber\\
        &\times \int \frac{d\Omega_3}{4\pi}  F  \frac{ \sum_{\rm spins}|\mathcal{M}_{\rm el}|^2}{\left|1-\frac{|\mathbf{p}_1+\mathbf{p}_2|\cos\theta_3}{2p_3}\right|}
    \end{align}
where, $\cos\theta_3=\hat{p}_1.\hat{p}_3$,  in going to the second line we took the non-relativistic limit $(E_i\approx m_\chi)$, and $p_3$ is a function of $\mathbf{p_1}$, $\mathbf{p_2}$, and $\cos\theta_3$ which is defined implicitly by the energy conservation equation
\begin{align}
    E_\chi(p_1)+E_\chi(p_2)=E_\chi(p_3)+E_(p_4)
\end{align}
with $E_\chi(p)=\sqrt{p^2+m_\chi^2}$ and 
$p_4=\sqrt{\left|\mathbf{p}_1+\mathbf{p}_2\right|^2-2|\mathbf{p}_1+\mathbf{p}_2|p_3\cos\theta_3+p_3^2}$. Noting that $\int_{p_i<\tilde{p}_F(\hat{p}_i)} d^3p_i/(2\pi)^3=n_\chi$ and $p_i\sim p_F$ for degenerate fermions, $\theta_3\sim 1$ for collisions that are not Pauli blocked (for which $F=1$), and assuming Pauli blocking amounts to an $\mathcal{O}(1)$ reduction in the final phase space, our best estimate for the typical elastic scattering rate is 
\begin{align}
    \Gamma_{\rm el}\sim \frac{n_\chi p_F\sum_{\rm spins}|\mathcal{M}_{\rm el}|^2}{8\pi(2m_\chi)^3}   \label{eq:Gammaelest}
\end{align}

The elastic scattering amplitude reads
\begin{align}
     \sum_{\rm spins}\left|\mathcal{M}_{\rm el}\right|^2=&g_S^4\left[\frac{\left(4m_\chi^2-t\right)^2}{\left(t-m_S^2\right)^2}+\frac{\left(4m_\chi^2-u\right)^2}{\left(u-m_S^2\right)^2}\right.\nonumber\\
    &\left.+\frac{t u-4m_\chi^2 s}{\left(t-m_S^2\right)\left(u-m_S^2\right)}\right]
\end{align}
To express the amplitude more explicitly, we move to the center-of-mass frame where the 3-momenta are related as $\mathbf{p_2^\prime}=-\mathbf{p_1^\prime}$ and $\mathbf{p_4^\prime}=-\mathbf{p_3^\prime}$. Next, we define 
$\cos\theta^\prime\equiv\hat{p}_1^\prime.\hat{p}_3^\prime=\hat{p}_1^\prime.\hat{p}_4^\prime$, $p^\prime\equiv p_1^\prime=p_2^\prime=p_3^\prime=p_4^\prime$, and $\eta^\prime=2p^{\prime 2}/m_S^2$, and then take the non-relativistic limit of the scattering amplitude
\begin{widetext}
\begin{align}
     \sum_{\rm spins}\left|\mathcal{M}_{\rm el}\right|^2=&\frac{16g_S^4m_\chi^4}{m_S^4}\left\{\frac{1}{\left[1+\eta'(1-\cos\theta')\right]^2}+\frac{1}{\left[1+\eta'(1+\cos\theta')\right]^2}-\frac{1}{\left[1+\eta'(1-\cos\theta')\right]\left[1+\eta'(1+\cos\theta')\right]}\right\}
\end{align}
\end{widetext}
A typical Pauli-allowed scattering has $\theta'\sim 1$ and $\eta'\sim 2p_F^2/m_S^2\gg 1$ (since $p_F\gtrsim E_F$ and we assume $E_F\gg m_S$). Thus, a good estimate for the spin-summed elastic scattering amplitude is 
\begin{align}
    \sum_{\rm spin} |\mathcal{M}_{\rm el}|^2\sim \frac{4g_S^4m_\chi^4}{p_F^4}\label{eq:Melsqest} 
\end{align}

Based on our best estimate of $\Gamma_{\rm el}$ as per Eqs.~\eqref{eq:Gammaelest} and \eqref{eq:Melsqest}, we write the exact elastic scattering rate as
\begin{align}
    \Gamma_{\rm el}(\mathbf{p}_1)&= \frac{c_{\rm el}(\mathbf{p}_1)}{8\pi} \times  \frac{g_S^4 m_\chi}{6\pi^2}
\end{align}
where we have used $n_\chi=p_F^3/(3\pi^2)$ and defined the following numerical factor that should be  $\mathcal{O}(1)$
\begin{align}
    c_{\rm el}(\mathbf{p}_1)=&\left(\frac{g_S^4 m_\chi}{6\pi^2}\right)^{-1}\frac{1}{(2m_\chi)^3}\int_{p_2<\tilde{p}_F(\hat{p}_2)}\frac{d^3p_2}{(2\pi)^3}p_3\nonumber\\
    &\times \int \frac{d\Omega_3}{4\pi}  F\frac{\sum_{\rm spins}|\mathcal{M}_{\rm el}|^2}{\left|1-\frac{|\mathbf{p}_1+\mathbf{p}_2|\cos\theta_3}{2p_3}\right|}
\end{align}
The above integral can be computed numerically if the distribution function of the fermions $\chi$ (and hence F) is known, perhaps through quantum Boltzmann simulations. We leave such a detailed study for future work and simply set $c_{\rm el}(\mathbf{p}_1)=1$, yielding the expression used in Appendix.~\ref{appendix:modeldetails}, Eq.~\eqref{eq:Gammaelmain}.

\subsection{Bremsstrahlung $\chi\chi\rightarrow\chi\chi S$}
\label{appendix:bremsstrahlung}
In terms of previously defined quantities, the luminosity of the heavy scalar $S$ emitted through the bremsstrahlung process $\chi(P_1)\chi(P_2)\rightarrow\chi(P_3)\chi(P_4)S(K)$ can be calculated as follows \cite{peskin2018introduction}
\begin{align}
    \frac{dL_S}{d\ln k}=&\int dV\frac{1}{2}\frac{k^3}{(2\pi)^32}\int d\Omega_S\int d\Pi_\chi F\left|\mathcal{M}_{\rm brem}\right|^2\nonumber\\
    &\times(2\pi)^4\delta^{(4)}\left(P_1+P_2-P_3-P_4-K\right)
\end{align}
where the newly defined quantities are the 4-momentum $K=(\omega,k)$ of the emitted heavy scalar $S$, $\int d\Pi_\chi=\int_{p_1<\tilde{p}_F(\hat{p}_1)} d\Pi_1\int_{p_2<\tilde{p}_F(\hat{p}_2)} d\Pi_2 \int d\Pi_3 d\Pi_4$, and the bremsstrahlung amplitude $\mathcal{M}$. In the soft radiation limit, $k\ll p_i$, the bremsstrahlung amplitude factorizes into an almost-elastic scattering part $\sum_{\rm spin}\left|\mathcal{M}_{\rm el}\right|^2$ and a part characterizing the probability of $S$ emission in a single collision $\mathcal{M}_S^2$ as
\begin{align}
    \sum_{\rm spins}
 \left|\mathcal{M}_{\rm brem}\right|^2=\mathcal{M}_S^2\times \sum_{\rm spin}\left|\mathcal{M}_{\rm el}\right|^2
\end{align}
with
\begin{align}
    \mathcal{M}_S&=g_Sm_\chi\left[\frac{1}{P_1.K}+\frac{1}{P_2.K}-\frac{1}{P_3.K}-\frac{1}{P_4.K}\right]\nonumber\\
    &=\frac{g_S\left[\left(\mathbf{p_1^\prime}.\mathbf{k^\prime}\right)^2+\left(\mathbf{p_2^\prime}.\mathbf{k^\prime}\right)^2-\left(\mathbf{p_3^\prime}.\mathbf{k^\prime}\right)^2-\left(\mathbf{p_4^\prime}.\mathbf{k^\prime}\right)^2\right]}{2m_\chi^2\omega^{\prime 3}}
\end{align}
where in going to the second line we took the non-relativistic $\chi$ limit. Furthermore, since $k\approx \omega\lesssim E_F\ll p_f$, we can safely neglect $\vec{k}$ in the 3-momentum conserving delta function (but still include $\omega$ in the energy one). After evaluating the $\int d\Pi_4$ and $\int dp_3$ integrals, the bremsstrahlung rate thus simplifies as 
\begin{widetext}
    \begin{align}
    \frac{dL_S}{d\ln k}=\int dV\frac{1}{2}\frac{k^3}{(2\pi)^3 2}\frac{1}{4\pi (2m_\chi)^3}\int_{p_1<\tilde{p}_F(\hat{p}_1)}\frac{dp_1^3}{(2\pi)^3}\int_{p_2<\tilde{p}_F(\hat{p}_2)}\frac{dp_2^3}{(2\pi)^3} p_3 \int \frac{d\Omega_3}{4\pi}  F  \frac{ \sum_{\rm spins}|\mathcal{M}_{\rm el}|^2}{\left|1-\frac{|\mathbf{p}_1+\mathbf{p}_2|\cos\theta_3}{2p_3}\right|}\int d\Omega_S \mathcal{M}_S^2\label{eq:dLSdlnkfull}
\end{align}
\end{widetext}
The $\int d\Omega_S \mathcal{M}_S^2$ can be computed in the center-of-mass frame in the non-relativistic limit ($p_i\ll m_\chi$)
\begin{align}
    \int d\Omega_S \mathcal{M}_S^2&\approx \frac{\pi g_S^2 k^4 (|\mathbf{p_1}-\mathbf{p_2}|^4)}{15m_\chi^4\omega^6}\left(1-\cos^2\theta'\right)\nonumber\\
    &\sim \frac{16\pi g_S^2k^4p_F^4}{15m_\chi^4\omega^6}
\end{align}
Based on the above estimate, $p_i\sim p_F$ and $\theta_3\sim 1$ for a typical Pauli-allowed collision, and the previously found estimate for $\sum_{\rm spins}|\mathcal{M}_{\rm el}|^2$ in Eq.~\eqref{eq:Melsqest}, we can again write $dL_S/d\ln k$ as an $\mathcal{O}(1)$ numerical factor $c_{\rm brem}$ times our best estimate for it
\begin{align}
    \frac{dL_S}{d\ln k}=c_{\rm brem}\times \frac{8N_\chi g_S^6p_F^4}{(2\pi)^545 m_\chi^3}\left(\frac{k}{\omega}\right)^6k
\end{align}
where
\begin{widetext}
\begin{align}
    c_{\rm brem}=\frac{\int dV\frac{1}{2}\frac{k^3}{(2\pi)^3 2}\frac{1}{4\pi (2m_\chi)^3}\int_{p_1<\tilde{p}_F(\hat{p}_1)}\frac{dp_1^3}{(2\pi)^3}\int_{p_2<\tilde{p}_F(\hat{p}_2)}\frac{dp_2^3}{(2\pi)^3} p_3 \int \frac{d\Omega_3}{4\pi}  F  \frac{ \sum_{\rm spins}|\mathcal{M}_{\rm el}|^2}{\left|1-\frac{|\mathbf{p}_1+\mathbf{p}_2|\cos\theta_3}{2p_3}\right|}\int d\Omega_S \mathcal{M}_S^2}{\frac{8N_\chi g_S^6p_F^4}{(2\pi)^545 m_\chi^3} \left(\frac{k}{\omega}\right)^6k}
\end{align}
\end{widetext}
and $2N_\chi=\int dV n_\chi$. More precisely, here $n_\chi$ is the total number density of $\chi$ particles (of both blobs combined) at an arbitrary moment during a blob collision, while $N_\chi$ is the total number of $\chi$ particles in a single, pre-collision blob. We have thus far been working in the soft radiation limit ($\omega\ll E_F$). To arrive at Eq.~\eqref{eq:dLSdlnw}, we set $\omega\approx k$ and take $c_{\rm brem}=1$.

\bibliography{references}
\bibliographystyle{apsrev4-1}

\end{document}